\documentclass[prl,twocolumn,preprintnumbers,footinbib,tightenlines,superscriptaddress]{revtex4}

\pdfoutput=1
\usepackage[english]{babel}
\usepackage{amsmath,amssymb,amsfonts, bm,bbm,slashed, subdepth}
\usepackage{graphicx}
\usepackage[sort&compress]{natbib}
\usepackage{xcolor}
\usepackage[normalem]{ulem}
\usepackage{hyperref}
\usepackage{cleveref}
\definecolor{red}{rgb}{1.0, 0, 0}
\usepackage{enumerate}
\usepackage{epsfig, subfigure}
\usepackage{setspace}
\usepackage{booktabs, tabularx}
\usepackage{units}
\usepackage{placeins}
\usepackage{multirow}
\usepackage{mathtools}

\newcommand{\res}{R}

\newcommand{\DM}{\text{DM}}

\begin{document}

\preprint{DESY 18-176}
\preprint{IPMU18-0161}
\title{Velocity Dependence from Resonant Self-Interacting Dark Matter}

\author{Xiaoyong Chu}
\email{xiaoyong.chu@oeaw.ac.at }
\affiliation{Institute of High Energy Physics, Austrian Academy of Sciences, Nikolsdorfer Gasse 18, 1050 Vienna, Austria
}

\author{Camilo Garcia-Cely}
\email{camilo.garcia.cely@desy.de}
\affiliation{Deutsches Elektronen-Synchrotron DESY, Notkestrasse 85,
22607 Hamburg, Germany}
\author{Hitoshi Murayama}
\email{hitoshi@berkeley.edu, hitoshi.murayama@ipmu.jp}
\affiliation{Department of Physics, University of California, Berkeley, CA 94720, USA}
\affiliation{Kavli Institute for the Physics and Mathematics of the
  Universe (WPI), University of Tokyo,
  Kashiwa 277-8583, Japan}
\affiliation{Ernest Orlando Lawrence Berkeley National Laboratory, Berkeley, CA 94720, USA}
\affiliation{Deutsches Elektronen-Synchrotron DESY, Notkestrasse 85,
22607 Hamburg, Germany}

\begin{abstract}
The dark matter density distribution in small-scale astrophysical objects may indicate that dark matter is  self-interacting,  while observations from clusters of galaxies suggest that the corresponding cross section depends on the velocity. 
Using a model-independent approach, we show  that resonant self-interacting dark matter (RSIDM) can naturally explain such a behavior. In contrast to what is often assumed,  this does not require a light mediator.  We  present explicit realizations of this mechanism and discuss the corresponding astrophysical constraints.

 \end{abstract}

\maketitle

Dark matter  (DM)  makes up more than 80\% of the matter in the Universe today and played a crucial role in forming stars and galaxies, and hence us.  Yet its nature is unknown.  Currently the best pieces of information come from astrophysical observations.
N-body simulations 
of collisionless DM predict astrophysical halos with DM density following a universal  profile that scales as $\rho \propto r^{-3}$ in its outskirts but   exhibits a central cusp, $\rho \propto r^{-\beta}$, with $\beta\simeq 1$,  referred to as the Navarro-Frenk-White (NFW) profile~\cite{Dubinski:1991bm, Navarro:1995iw,Navarro:1996gj}.
Nevertheless, many studies show hints of a DM mass deficit in the inner regions of  certain halos. Notably, observations indicate that numerous  dwarf galaxies~\cite{Moore:1994yx,Flores:1994gz, Walker:2011zu} and some  low-surface-brightness spiral galaxies~\cite{deBlok:2001hbg,deBlok:2002vgq,Simon:2004sr} have a shallower central DM density, better described by a core of constant density, {\it i.e.}\/, by $\beta\simeq 0$. 
This is known as the core-vs-cusp problem.
Although it is more pressing in small-scale objects, shallower DM density profiles --with a slope of $\beta\simeq 0.5$-- have been reported for certain galaxy clusters~\cite{Sand:2003bp,Newman:2012nw}.  
Moreover, the DM mass deficit also manifests itself in halos that are less dense than what simulations suggest if they host the galaxies that we observe. This is the too-big-to-fail problem, observed for the subhalos of the Milky Way~\cite{BoylanKolchin:2011de}, Andromeda~\cite{Tollerud:2014zha} and the Local Group~\cite{Kirby:2014sya}. 

\begin{figure*}[t]
\includegraphics[width=0.47\textwidth]{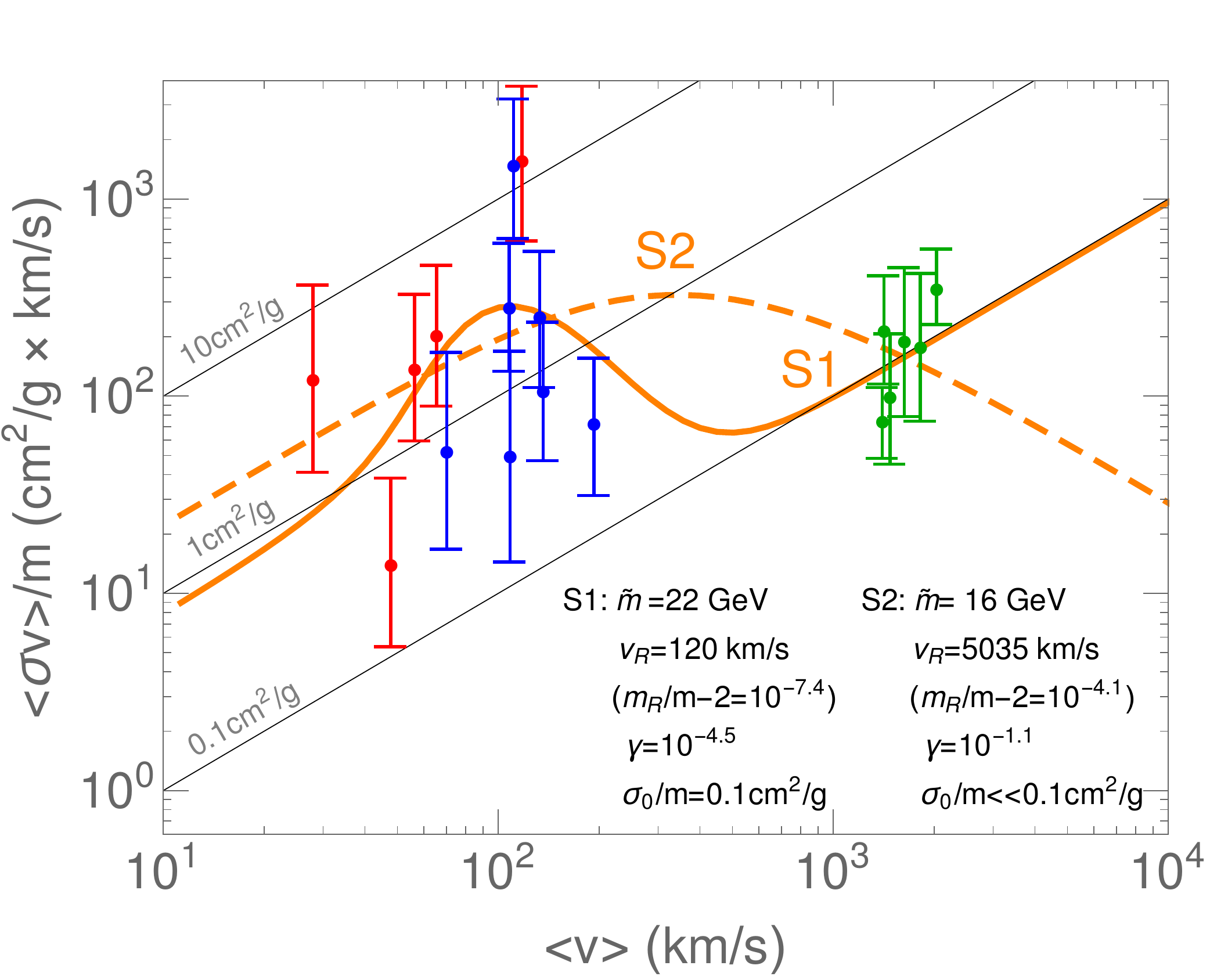}
\includegraphics[width=0.47\textwidth]{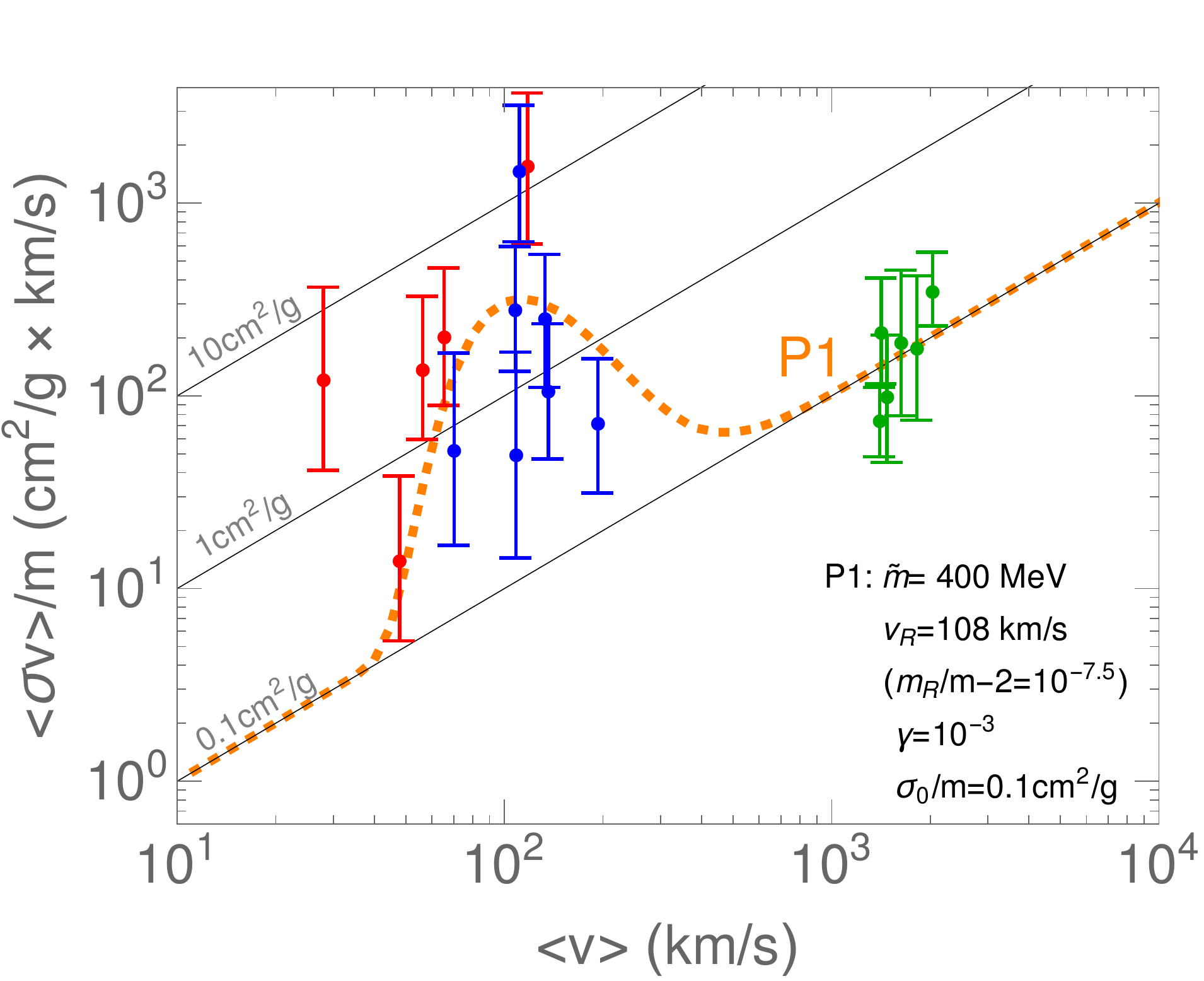}
\caption{ RSIDM cross section per unit of mass as a function of the velocity.  Best-fit curves to data \cite{Kaplinghat:2015aga} for $S$-wave (left) and $P$-wave scatterings (right). The latter is also the best-fit curve for $L>1$ after rescaling the mass with Eq.~\eqref{eq:pwavef2}. Here $\tilde{m} = m S^{-1/3}$. See text for details. }
\label{fig:sigmavsv}
\end{figure*}

Several explanations for these discrepancies have been discussed in the literature. The systematic uncertainties introduced in deriving DM distributions from observations of luminous objects are one of them.
Most importantly, the motions of HI gas and stars may not be faithful tracers of the DM circular velocity~\cite{Blok:2002tr, Rhee:2003vw, Gentile:2005de, Spekkens:2005ik,  Valenzuela:2005dh, Dalcanton:2010bp, Kormendy:2014ova, 2016MNRAS.462.3628R, Maccio:2016egb,  2017A&A...601A...1P, Brooks:2017rfe, Oman:2017vkl, 2018MNRAS.474.1398G, Read:2018pft}.  Baryonic processes are another conceivable explanation for the discrepancies, since the aforementioned simulations only include collisionless DM. Solutions along this line include supernova-driven baryonic winds~\cite{Navarro:1996bv,Gelato:1998hb, Binney:2000zt, Gnedin:2001ec}, DM heating due to star formation~\cite{Read:2018fxs}, infalling baryonic clumps~\cite{ElZant:2001re, Weinberg:2001gm, Ahn:2004xt, Tonini:2006gwz} as well as  active galactic nuclei or black holes~\cite{Martizzi:2011aa}. 
Nonetheless, there is no consensus on why systematic  uncertainties or baryonic processes lead to a seemingly universal mass deficit at various scales. 

A more exciting possibility consists of considering 
DM collisions in the inner regions of astrophysical objects~\cite{Spergel:1999mh}. This is known as self-interacting dark matter (SIDM). N-body simulations~\cite{Dave:2000ar,Vogelsberger:2012ku,Rocha:2012jg,Peter:2012jh,Elbert:2014bma,Fry:2015rta} confirm that DM scattering processes  indeed reduce the central density of DM halos,  
providing a solution to  both problems ~\footnote{Besides, SIDM can also explain the diversity of galaxy rotation curves~\cite{Oman:2015xda,Kamada:2016euw,Creasey:2016jaq,Robertson:2017mgj}.}. For a recent review see~\cite{Tulin:2017ara}.

The observed mass deficit is more appreciable in small-scale halos, where the DM velocity dispersion is relatively low. Therefore, a self-scattering cross section that decreases with the DM velocity  can better fit observations~\cite{Kaplinghat:2015aga}, although a constant cross section is certainly not excluded due to the large uncertainties mentioned above. 
A long-range force induced by a light boson interacting with DM is often invoked to obtain a velocity-dependent  cross section~\cite{Spergel:1999mh, Feng:2009hw}. Other possibilities that do not involve a light mediator  include exothermic inelastic scatterings~\cite{McDermott:2017vyk, Vogelsberger:2018bok} and self-heating DM~\cite{Kamada:2017gfc, Chu:2018nki, Kamada:2018hte}. 

The essence of this work is to discuss the resonant self-interaction of DM (RSIDM) as another mechanism  for achieving the desired  velocity dependence of SIDM. Such a resonant behavior  was firstly discussed for DM annihilation in \cite{Griest:1990kh, Gondolo:1990dk, Jungman:1995df, Feldman:2008xs, Pospelov:2008jd, Ibe:2008ye,  MarchRussell:2008tu, Guo:2009aj, Ibe:2009dx,Kakizaki:2005en,Arina:2014fna}, and
 applied to DM self-scattering in specific scenarios~\cite{Ibe:2009mk,Braaten:2013tza, Duch:2017nbe, Braaten:2018xuw}.  Nevertheless, the velocity dependence of resonant self-scattering and its general astrophysical consequences  have not been explored in detail. In this letter,  we do so    in a model-independent way, and show that  
resonant scattering is able to address the observed DM mass deficit at all astrophysical scales.  Concrete DM scenarios and indirect searches are discussed later.

{\bf Resonant scattering in DM halos.}
Numerous studies claim that the density distribution of certain DM halos do not  follow a NFW profile in the inner region.  In the SIDM hypothesis, this is due to DM collisions that thermalize the DM particles in such a region thereby reducing its average density~\cite{Spergel:1999mh}. Hence the inner profile is closely related to the velocity-averaged scattering cross section per unit of DM mass,    $\langle \sigma v\rangle/m$, where
\footnote{
For resonant scattering, $\sigma$ nearly equals  the momentum-transfer cross section, $\sigma_T $, and we do not differentiate between  them here.}
\begin{equation}
\langle \sigma v \rangle = \int^{v_\text{max}}_0 f(v,v_0) \sigma v dv \,,\quad f (v,v_0) = \frac{4 v^2e^{-{v^2}/{v_0^2}}}{\sqrt{\pi}v_0^3} \,.
\end{equation}
Here, $v$ is the relative velocity, which we assume to follow  a Maxwell-Boltzmann distribution truncated at the escape velocity,  $v_\text{max}$, of the corresponding halo. $v_0$ is a parameter related to the average relative velocity via $\langle v\rangle \simeq 2 v_0/ \sqrt{\pi}$. Notice that in dwarf galaxies $\langle v \rangle \sim 20$ km/s whereas in clusters of galaxies $\langle v \rangle \sim 2000$ km/s.

A semi-analytical method has been proposed in \cite{Kaplinghat:2015aga} to infer the   value of $\langle \sigma v\rangle/m$ for a given DM halo from  observational data. The  method was applied to five clusters from \cite{Newman:2012nw},    seven low-surface-brightness spiral galaxies  in \cite{KuziodeNaray:2007qi} and six  dwarf galaxies of the THINGS sample~\cite{Oh:2010ea} (also see \cite{Elbert:2016dbb, Valli:2017ktb}). Fig.~\ref{fig:sigmavsv} shows their results in green,  blue and red, respectively. The values presented here are for illustrative purpose, and should be taken with caution due to the large uncertainties in extracting the cross sections from kinematical data. See e.g. \cite{sokolenko:2018noz} for a recent study.
Nonetheless, at face value, the figure demonstrates that a cross section independent of the velocity --the ones corresponding to the diagonal lines-- can hardly accommodate all points. 
Notice that the values of $\sigma/m$ at cluster scales  
are in agreement with  observations from the Bullet Cluster giving $\sigma/m\lesssim\unit[1.3]{cm^2/g}$~\cite{Randall:2007ph,Robertson:2016xjh}, which is one of the strongest constraints on DM self-interactions.

Barring the uncertainties, the figure suggests that the cross section depends on $\langle v\rangle$. In this letter, we propose that this is due to RSIDM.  This takes place when there exists an intermediate particle, denoted as $R$, so that the total self-scattering cross section can be cast as a sum of a constant piece, $\sigma_0$, plus a Breit-Wigner resonance\,\footnote{Note that the interference term only exists for $S$-wave scattering. For the cases discussed here, that term was found to be negligible with respect to the second term of Eq.~\eqref{eq:BW}. Furthermore,  it changes its sign from below to above resonance and hence nearly cancels out upon integration over the velocity profile.}. More explicitly, for non-relativistic DM,
\begin{eqnarray}
\label{eq:res}
\sigma&=&\sigma_0+\frac{4\pi\,S}{m E(v)}\cdot \frac{\Gamma(v)^2/4 }{\left( E(v)- E(v_R)\right)^2+ \Gamma(v)^2/4 }\,,
\label{eq:BW}
\end{eqnarray}
where  the total kinetic energy and symmetry factor read
\begin{align}
E(v) = \frac{1}{2}\frac{m}{2} v^2 &&\text{and}&&
S= \frac{  2 J_R+1}{ (2 J_\DM+1)^2 }\,.
\end{align}
Here, $J_R$ and $J_\DM$ are the spins of the resonance and the DM particles, respectively.  $m/2$ is the reduced mass.  If DM has internal degrees of freedom other than its spin, they must be accounted for in $S$. 
The collision hits the resonance when $v= v_R$ and hence $E(v_R)= m_R-2m$. 

\begin{figure}[t]
\includegraphics[trim=1.2cm 0cm 0cm 0cm,clip,width=0.54\textwidth]{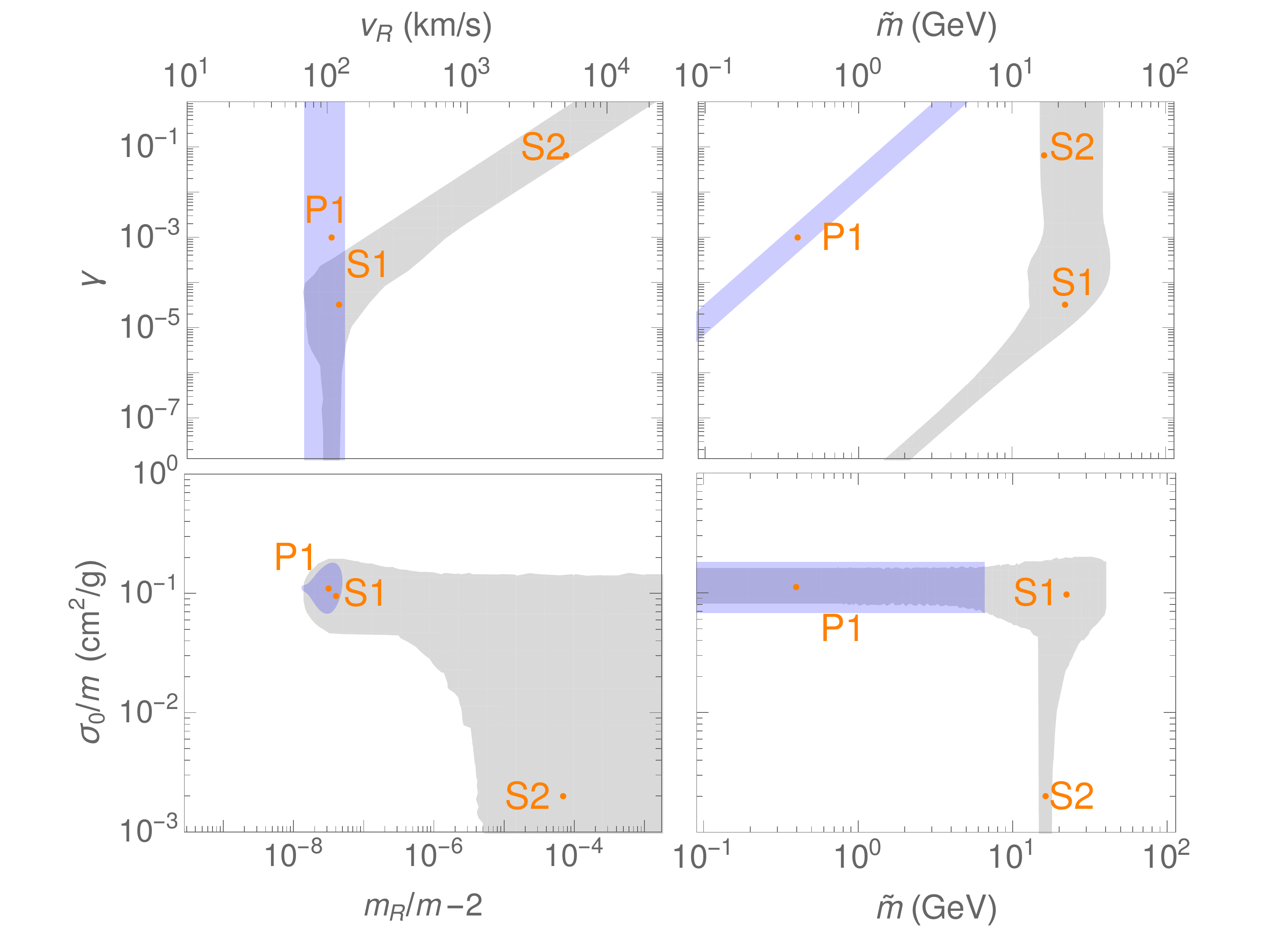}
\caption{ 95\% C.L. contours for $S$-wave (gray) and $P$-wave (purple) scatterings together with the corresponding parameter sets of Fig.~\ref{fig:sigmavsv}. Notice that $m_R/m-2=v_R^2/4$.}
\label{fig:fit}
\end{figure}

In addition, the width in Eq.~\eqref{eq:BW}  can be calculated in terms of the resonance self-energy by means of  $\Gamma(v) = {\text{Im}} \, \Sigma (v)/m_R$. This, as well as the denominator  in Eq.~\eqref{eq:BW}, assumes that the total width is dominated by the process $R\to\text{DM\,DM}$. Besides that, Eq.~\eqref{eq:BW} is completely general as it directly follows from unitarity  considerations of  the scattering matrix~\cite{Breit:1936zzb}. In perturbative theories, the running width can be written as~\footnote{
This can be proven using effective range theory~\cite{Bethe:1949yr}.}
\begin{align}
\Gamma(v)= m_R \gamma v^{2L+1}\,.
\label{eq:width}
\end{align}
Here, $L$ is the orbital angular momentum, $\Gamma(v_R)$ is the decay rate, and a  constant $\gamma\lesssim O(1)$  characterizes the coupling between the resonance and DM. The factor $v^{2L+1}$ accounts for the phase space and possible angular momentum suppression.  
Then we find $\langle\sigma v \rangle = \sigma_0\langle v \rangle+ 256 \pi\, S \, {\cal I}_L (\gamma, v_R,v_0)/m^2$, where a dimensionless 
\begin{equation}
{\cal I}_L (\gamma, v_R,v_0)\equiv
\int^{v_\text{max}}_0  \frac{\gamma^2 f(v,v_0) v^{4L+1}\, dv}{(v^2-v_R^2)^2+ 16\gamma^2 v^{2(2L+1)}}\,
\label{eq:IL}
\end{equation}
 determines the non-trivial velocity-dependence of the resonant self-scattering.  For $S$-wave and $P$-wave scatterings, we calculate the best-fit parameter sets S1, S2, and P1 based on the inferred data from Ref.~\cite{Kaplinghat:2015aga}  and show them in Fig.~\ref{fig:sigmavsv}~\footnote{
Using a similar method, $\langle\sigma v \rangle/m$ has been estimated for eight Milky Way dwarfs~\cite{ Valli:2017ktb}.    A proper combined fit including those results is beyond the scope of this letter. Nevertheless, since such dwarf galaxies have all approximately the same $\langle v \rangle$, a combined fit would not change our conclusions regarding the velocity dependence of $\langle\sigma v \rangle$.
}.  $\sigma_0/m$ is fitted with the other parameters for S1 and P1 while  for S2 a negligible $\sigma_0/m \ll \unit[0.1]{cm^2/g}$ is taken as a prior. 
They all lead to $\chi^2/\text{d.o.f.}\simeq 2$, in contrast to $\chi^2/\text{d.o.f.}\simeq 6$ for the fit assuming only a constant cross section (we treat errors as uncorrelated). For S1 and P1, we show the 95\% C.L. contours in Fig.~\ref{fig:fit}. 
Many comments are in order. 

 First, we have numerically checked that a precise knowledge of the escape velocity is not necessary for calculating ${\cal I}_L$.  
This is because Eq.~\eqref{eq:IL} converges quite fast due to  
the Boltzmann  
factor. In fact, as shown in the supplementary material, exact solutions exist in the limit $v_\text{max}\gg v_0$, which will be implicitly applied  hereafter for simplicity.

Second, to qualitatively understand Figs.~\ref{fig:sigmavsv} and ~\ref{fig:fit}, one can use the narrow-width approximation (NWA)
\begin{equation}
\frac{1}{(v^2-v_R^2)^2+16\gamma^2 v^{2(2L+1)}}\to \frac{\pi \delta(v-v_R)}{8\gamma v_R^{2(L+1)}}\,.
\label{eq:swa}
\end{equation} 
It works very well  for $L\ge 1 $ because $\gamma^2 v^{2(2L+1)} \ll v^4$. 
In this case, we find that ${\cal I}_L (\gamma, v_R,v_0)$ scales as $\gamma^2 v_0^{4L+1}/v_R^4 $ at $v_0\ll v_R$, and  as $\gamma^2 v_0^{4L-3} $ at $v_0\gg v_R$.  In both regions, ${\cal I}_L$ can not be much larger than one. Therefore, the resonant effect is negligible except for the intermediate region, where the NWA captures the velocity dependence as
\begin{equation}
\frac{\langle\sigma v \rangle }{m} \bigg|_\text{NWA} =\frac{\sigma_0\langle  v \rangle }{m}+ \frac{128 S\pi^{3\over 2}  \gamma  v_R^{2L+ 1} }{m^3 v_0^3} e^{-{v_R^2}/{v_0^2}} \,.
\label{pwave:IL}
\end{equation}
Notice that the peak lies at $v_0 \sim v_R $ as illustrated by P1 in Fig.~\ref{fig:sigmavsv}. The corresponding line   actually applies to any $L\ge 1$, because the dependence on $L$  can be absorbed by rescaling $m$.  Using Eq.~\eqref{pwave:IL} we find that the best-fit parameters  at 95\% C.L. for $L\geq 1$ are given by
\begin{eqnarray}
&v_R = \unit[108^{+28}_{-43} ]{km/s}\,,\qquad \sigma_0/m = \unit[0.11^{+0.10}_{-0.05} ]{cm^2/g}\,,\label{eq:pwavef}\\
&\tilde{m} = \unit[400^{+120}_{-90}]{MeV}\cdot  \left(\frac{\gamma}{10^{-3}}\right)^{\frac{1}{3}}\left(\frac{v_R}{\unit[3\cdot 10^5]{km/s} }\right)^{{2(L-1)}/{3}}.
\label{eq:pwavef2}
\end{eqnarray}
Such values for the velocity correspond to $m_R/m-2 \sim 10^{-7}$.   
The regions where all this applies are shown in  Fig.~\ref{fig:fit}. For $P$-wave scattering, demanding $\gamma \lesssim 1$ leads to $\tilde{m}\equiv m \,S^{-1/3}\lesssim\unit[5]{GeV}$. Moreover, a perturbative $\sigma_0/m$ around $0.1$\,cm$^2/$g  requires sub-GeV DM masses unless $S\gg 1$.
Interestingly, P1 predicts $\sigma/m\sim\unit[0.1]{cm^2/g}$ at $\langle v \rangle \ll\unit[100]{km/s}$. In fact, scatterings with $L\geq 1$ can realize small cross sections  at very low velocities. Hence,  the recent claim based on Draco observations~\cite{Read:2018pft} is consistent with RSIDM.

As long as $v_R \gtrsim  4\gamma$, the NWA also applies for $S$-wave scattering. 
For  $v_R \ll  4\gamma$,  
${\cal I}_L$ is proportional to  $v_0$  (to $1/v_0$) below (above)   $v_\text{peak} \sim v_R^2/(4\gamma) \ll v_R$,  because such large values of $\gamma$ broaden the resonance.   
 S1 and S2 illustrate the narrow and the broad width cases, respectively.

In conclusion, resonant scattering is able to address the observed DM mass deficit at all astrophysical scales. 

\begin{table}
\begin{tabular}{llccccc}\hline
Scenario & Interaction Lagrangian& $L$& $J_\DM$ & $J_R^{~P}$ & $S$ & $\gamma$\\\hline
I &$  g\, \res \,\overline{\DM}\gamma^5 \DM $& 0 &  $\frac{1}{2}$ & 0$^-$  & $\frac{1}{4}$ & $\frac{ g^2}{32 \pi}$\\
IIa &  $ \,g R \, \DM^i  \DM^i $
& 0 &  0 & 0$^+$ &  $\frac{1}{3}$  
 & $ \frac{g^2}{16 \pi \,m_R^2}$\\
IIb &  $g \,\epsilon_{ijk}\, R_\mu^i \, \DM^j \partial^\mu \DM^k $
& 1 &  0 & 1$^-$ &  $ 1$ 
& $ \frac{g^2}{384 \pi}$\\
III & $  \frac{1}{\Lambda} R_{\mu \nu }\, {\cal T}_\DM^{\mu\nu} $& 2 & 0 & 2$^+$ &  5 & $\frac{m_R^2}{30720 \pi \Lambda^2}$
\end{tabular}
\caption{Benchmark RSIDM models.  
}
\label{table}
\end{table}

{\bf RSIDM Models.}
Below we illustrate the previous model-independent results in concrete RSIDM scenarios. We first introduce a Lagrangian specifying the coupling of the DM to the resonance (see Table~\ref{table}) and calculate the cross section and the self-energy. We subsequently corroborate that they can be cast as Eqs.~\eqref{eq:BW} and \eqref{eq:width} show. The scenarios are:

I. \textit{Fermionic DM  with a pseudoscalar mediator. }
The scattering process is $S$-wave while  
$\sigma_0 \simeq 0$.  The corresponding best fit is thus S2. Notice that a light pseudoscalar mediator does not lead to SIDM  because it induces a suppressed Yukawa potential (see e.g.~\cite{Kahlhoefer:2017umn}).  Due to this and because it leads to velocity-suppressed direct-detection rates, this candidate is phenomenologically interesting.

II. \textit{Dark mesons}. In QCD-like theories, DM can be a dark pion.  Analogous to real pions, it can be a triplet $\DM^{i}$, with $i=1,2,3$. If $R$ is a dark $\sigma$ resonance (IIa), the scattering takes place via the $S$-wave, where we expect GeV DM and $\sigma_0/m \ll \unit[0.1]{cm^2}/g$. The best fit is thus S2. If $R$ is a dark $\rho$ resonance (IIb), the scattering is $P$-wave suppressed. The constant piece of the cross section is given by $ \sigma_0 \sim \pi\gamma^2/ m^2$ in perturbation theory, but it is plausible that there are other contributions. We therefore leave $\sigma_0$ as a free parameter. The corresponding best-fit curve is P1. We expect $m \sim \unit[400]{MeV}$ in this case. In the same fashion, minimal QCD-like theories can also lead to spin-1 DM~\cite{Francis:2018xjd}.  In all cases, DM can be produced by means of the SIMP \cite{Dolgov:1980uu,Carlson:1992fn,Hochberg:2014dra,Yamanaka:2014pva,Hochberg:2014kqa,Bernal:2015bla,Bernal:2015lbl,Lee:2015gsa,Choi:2015bya,Hansen:2015yaa,Bernal:2015xba, Kuflik:2015isi,Hochberg:2015vrg,Choi:2016hid,Pappadopulo:2016pkp,Farina:2016llk,Choi:2016tkj,Dey:2016qgf,Cline:2017tka, Choi:2017mkk, Choi:2017zww,Chu:2017msm, Choi:2018iit}   and the freeze-in~\cite{McDonald:2001vt,Hall:2009bx,Bernal:2015ova} mechanisms.

III.  \textit{Tensor resonances}. They also arise in strongly-coupled theories. Despite the potential complications of such theories, the generality of our approach allows to describe the scattering induced by a spin-2 resonance $R_{\mu\nu}$~\footnote{Not to be confused with the Ricci tensor.}. If this couples to the DM energy-momentum tensor with a cut-off scale $\Lambda$, and  taking scalar DM as an example, we find that the corresponding Feynman rules~\cite{Han:1998sg} indeed lead to a $D$-wave cross section  given by Eq.~\eqref{eq:BW}.  For $m\sim 10^{-3} \Lambda$, we obtain keV DM with $\gamma \sim 10^{-13}$.  The corresponding best fit is given by P1 in Fig.~\ref{fig:sigmavsv} after  rescaling the mass by means of Eq.~\eqref{eq:pwavef2}.

{\bf {Annihilation vs. Scattering.}}
It is not necessary that the DM annihilates, as \textit{e.g.} in models of asymmetric DM. 
Nonetheless, if the resonance decays into a pair of Standard Model (SM) particles $f\overline{f}$, in analogy to Eq.~\eqref{eq:BW}, the resonant DM annihilation into  $f\overline{f}$ has a cross section
 \begin{eqnarray}
\label{eq:resSM}
\sigma_\text{anni}&\simeq&\frac{4\pi\,S}{m E(v)}\cdot \frac{ \Gamma(v) \cdot m_R\gamma_f/4 }{\left( E(v)- E(v_R)\right)^2+ \Gamma(v)^2/4 }\,,
\end{eqnarray} 
where  $m_R\gamma_f$ is the decay width for $R\to f\overline{f}$. As above, we assume that the resonance dominantly decays to a pair of DM particles, and thus that the contribution of $f$ to the imaginary part of the resonance self-energy, $\,m_R^2 \gamma_f$, is subleading.  This is different from \cite{Ibe:2009mk}, in which the resonance dominantly decays into visible particles.
As expected for annihilations (but not for elastic scatterings), $\sigma_\text{anni} v\propto v^{2L}$ as long as $v\ll v_R$. Furthermore, for the cases where NWA applies, $\langle \sigma_\text{anni} v\rangle\big|_\text{peak} \sim 32 \pi^2 S\, \gamma_f/(m^2 v_R^3) $.  In contrast, for broad $S$-wave resonances such as S2, where $v_\text{peak} \ll v_R$, $\langle \sigma_\text{anni} v\rangle\big|_\text{peak} $ gets enhanced by another factor $(v_R/v_\text{peak})^{2L+1}$. 
 
\begin{figure}
\includegraphics[width=0.49\textwidth]{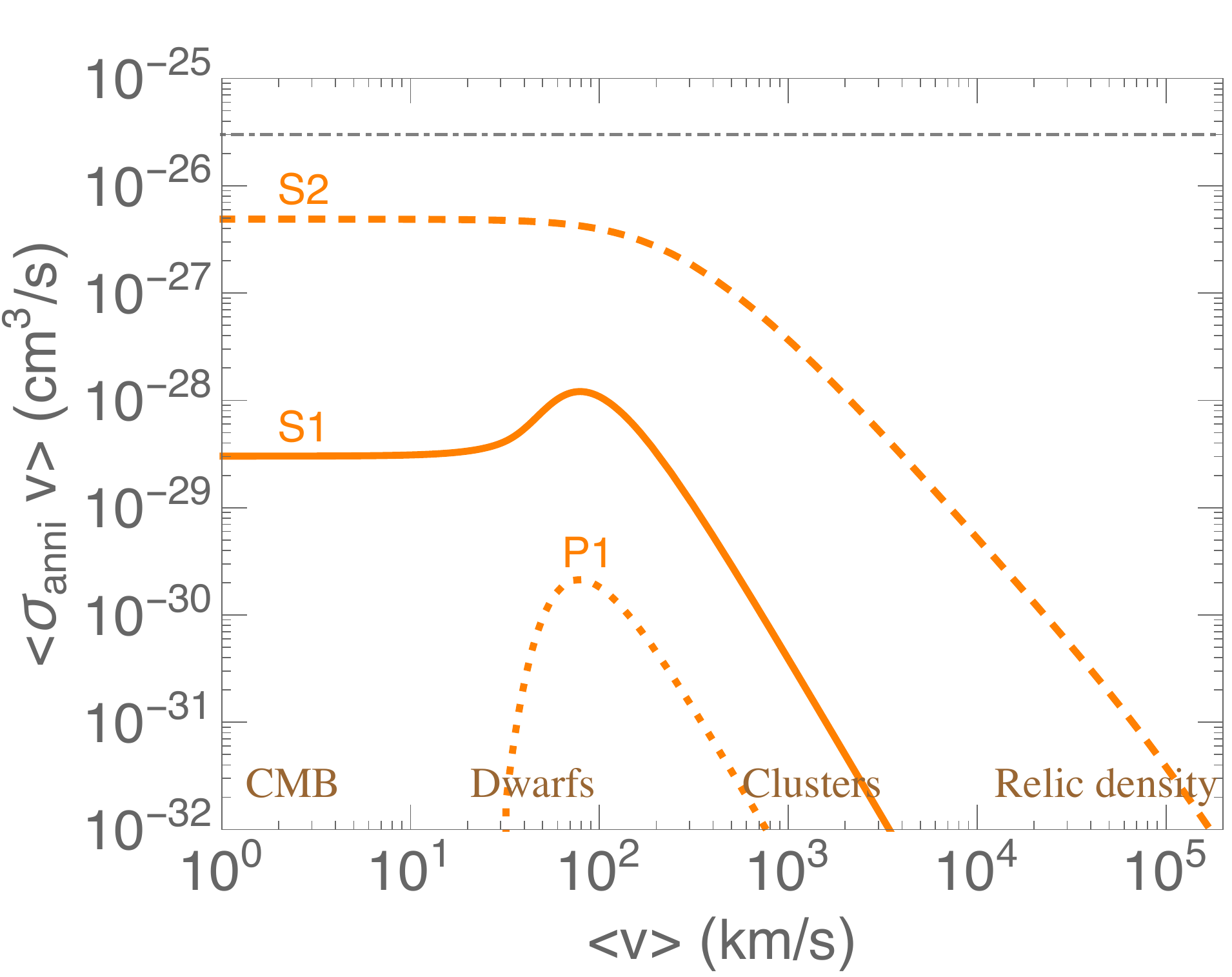}
\caption{Annihilation cross section into a pair of charged fermions for the parameter sets of Fig.~\ref{fig:sigmavsv}, assuming a branching ratio $\gamma_f/(\gamma v_R^{2L+1})= 10^{-13}$. The horizontal line gives the standard freeze-out benchmark.  }
\label{fig:annihilation}
\end{figure}

The coupling to light charged particles is mostly constrained by  Fermi-LAT observations of local satellites~\cite{Ackermann:2013yva, Fermi-LAT:2016uux} and the Planck data on the cosmic microwave background (CMB)~\cite{Ade:2015xua, Liu:2016cnk}.  
For instance, the corresponding Fermi-LAT  upper limit on $\langle \sigma v_\text{anni}\rangle$ for  
GeV DM  is of the order of   
$\unit[10^{-26}]{cm^3/s}$. For S2, this  leads to an upper limit on the branching ratio, $\gamma_f/(\gamma v_R)$, of about $ 10^{-13}$--$10^{-12}$. This bound is much stronger than that of S1 and P1, due to the enhancement factor mentioned above. 
Motivated by this, we conservatively fix  $\gamma_f/(\gamma v_R^{2L+1})= 10^{-13}$ and calculate the annihilation cross section as a function of $\langle v  \rangle$ for the same parameter sets of Fig.~\ref{fig:sigmavsv}. The result is shown in Fig.~\ref{fig:annihilation}.
Therefore, the resonance can only  couple feebly to light charged particles, which is why the SIDM candidates with thermal freeze-out from \cite{Duch:2017nbe} are excluded. Of course this is model-dependent. For instance, if  the resonance only couples to neutrinos, the bound on   $\langle \sigma_\text{anni} v\rangle$  becomes much weaker, and larger $\gamma_f/\gamma $ are thus allowed. 

Furthermore, the strong velocity dependence of $\langle\sigma v_\text{anni}\rangle$  suggests that the usual freeze-out can hardly work, as for SIDM with light mediators decaying into visible particles~\cite{Bringmann:2016din, Binder:2017lkj,Hufnagel:2017dgo,Hufnagel:2018bjp}. Nevertheless, the DM  abundance might arise from other SIDM production mechanisms~\cite{Bernal:2015ova}.
Indeed, for the $S$-wave case,  producing the DM abundance  with  small couplings is possible via  freeze-in~\cite{McDonald:2001vt,Hall:2009bx} or 4-to-2 annihilations~\cite{Bernal:2015xba},  where a scalar (vector) resonance can feebly mix with the Higgs (SM gauge bosons).
See \cite{Essig:2013lka, Bernal:2017kxu} for reviews.

{\bf Discussion.}
We advocate the resonant scattering  as a possible SIDM realization  with a velocity-dependent scattering cross section. Instead of a light mediator,  this RSIDM scenario requires a near-threshold resonance with $m_R/m-2$ ranging from $\sim 10^{-7}$ for narrow resonances to $10^{-2}$ for $S$-wave scattering with broad widths.
Such resonances exist in Nature. As an example, $\alpha$ particles resonantly scatter by means of $^{8}_4$Be in exactly the same way as described above. In fact, these processes were the main subject of the original article by Breit and Wigner~\cite{Breit:1936zzb} and they may as well occur in the DM sector. Actually, dark nucleons as SIDM have been studied in \cite{Braaten:2018xuw}~\footnote{Their findings suggest that their model is described by S2.}. Furthermore, lattice studies suggest that QCD-like theories of DM might possess such states~\cite{Briceno:2017max}.

{\bf Conclusions.} We find that this RSIDM hypothesis can certainly address the core-vs-cusp and the too-to-big-fail problems 
while still being in agreement with cluster observations.  
We have also discussed indirect detection signatures, which are nevertheless model-dependent. Additionally, we would like to emphasize that usual SIMPs --which are often said to be disfavored because their scattering cross section does not vary with velocity-- can easily accommodate the mechanism proposed here.

\begin{acknowledgements}
{\bf Acknowledgements.} We thank Ranjan Laha and Kai Schmidt-Hoberg for interesting discussions. X.C. is supported by the `New Frontiers' program of the Austrian Academy of Sciences. C.G.C. is supported by the ERC Starting Grant NewAve (638528).
 H.M. thanks the Alexander von Humboldt Foundation for support while this work was completed.  H.M. was supported by the NSF grant PHY-1638509, by the U.S. DOE Contract DE-AC02-05CH11231, by the JSPS Grant-in-Aid for Scientific Research (C) (17K05409), MEXT Grant-in-Aid for Scientific Research on Innovative Areas (15H05887, 15K21733), by WPI, MEXT, Japan, and  by the Binational Science Foundation  (grant No. 2016153).  
\end{acknowledgements}

\bibliographystyle{utcaps_mod}
\bibliography{ref}

\providecommand{\href}[2]{#2}\begingroup\raggedright\begin{thebibliography}{100}

\bibitem{Dubinski:1991bm}
J.~Dubinski and R.~G. Carlberg, ``{\em {The Structure of cold dark matter
  halos}},''
\href{http://dx.doi.org/10.1086/170451}{Astrophys. J. {\normalfont \bfseries
  378} (1991)  496}.

\bibitem{Navarro:1995iw}
J.~F. Navarro, C.~S. Frenk, and S.~D.~M. White, ``{\em {The Structure of cold
  dark matter halos}},'' \href{http://dx.doi.org/10.1086/177173}{Astrophys. J.
  {\normalfont \bfseries 462} (1996)  563--575},
\href{http://arxiv.org/abs/astro-ph/9508025}{{\normalfont \ttfamily
  arXiv:astro-ph/9508025}}.

\bibitem{Navarro:1996gj}
J.~F. Navarro, C.~S. Frenk, and S.~D.~M. White, ``{\em {A Universal density
  profile from hierarchical clustering}},''
  \href{http://dx.doi.org/10.1086/304888}{Astrophys. J. {\normalfont \bfseries
  490} (1997)  493--508},
\href{http://arxiv.org/abs/astro-ph/9611107}{{\normalfont \ttfamily
  arXiv:astro-ph/9611107}}.

\bibitem{Moore:1994yx}
B.~Moore, ``{\em {Evidence against dissipationless dark matter from
  observations of galaxy haloes}},''
\href{http://dx.doi.org/10.1038/370629a0}{Nature {\normalfont \bfseries 370}
  (1994)  629}.

\bibitem{Flores:1994gz}
R.~A. Flores and J.~R. Primack, ``{\em {Observational and theoretical
  constraints on singular dark matter halos}},''
  \href{http://dx.doi.org/10.1086/187350}{Astrophys. J. {\normalfont \bfseries
  427} (1994)  L1--4},
\href{http://arxiv.org/abs/astro-ph/9402004}{{\normalfont \ttfamily
  arXiv:astro-ph/9402004}}.

\bibitem{Walker:2011zu}
M.~G. Walker and J.~Penarrubia, ``{\em {A Method for Measuring (Slopes of) the
  Mass Profiles of Dwarf Spheroidal Galaxies}},''
  \href{http://dx.doi.org/10.1088/0004-637X/742/1/20}{Astrophys. J.
  {\normalfont \bfseries 742} (2011)  20},
\href{http://arxiv.org/abs/1108.2404}{{\normalfont \ttfamily arXiv:1108.2404}}.

\bibitem{deBlok:2001hbg}
W.~J.~G. de~Blok, S.~S. McGaugh, A.~Bosma, and V.~C. Rubin, ``{\em {Mass
  density profiles of LSB galaxies}},''
  \href{http://dx.doi.org/10.1086/320262}{Astrophys. J. {\normalfont \bfseries
  552} (2001)  L23--L26},
\href{http://arxiv.org/abs/astro-ph/0103102}{{\normalfont \ttfamily
  arXiv:astro-ph/0103102}}.

\bibitem{deBlok:2002vgq}
W.~J.~G. de~Blok and A.~Bosma, ``{\em {High-resolution rotation curves of low
  surface brightness galaxies}},''
  \href{http://dx.doi.org/10.1051/0004-6361:20020080}{Astron. Astrophys.
  {\normalfont \bfseries 385} (2002)  816},
\href{http://arxiv.org/abs/astro-ph/0201276}{{\normalfont \ttfamily
  arXiv:astro-ph/0201276}}.

\bibitem{Simon:2004sr}
J.~D. Simon, A.~D. Bolatto, A.~Leroy, L.~Blitz, and E.~L. Gates, ``{\em
  {High-resolution measurements of the halos of four dark matter-dominated
  galaxies: Deviations from a universal density profile}},''
  \href{http://dx.doi.org/10.1086/427684}{Astrophys. J. {\normalfont \bfseries
  621} (2005)  757--776},
\href{http://arxiv.org/abs/astro-ph/0412035}{{\normalfont \ttfamily
  arXiv:astro-ph/0412035}}.

\bibitem{Sand:2003bp}
D.~J. Sand, T.~Treu, G.~P. Smith, and R.~S. Ellis, ``{\em {The dark matter
  distribution in the central regions of galaxy clusters: Implications for
  CDM}},'' \href{http://dx.doi.org/10.1086/382146}{Astrophys. J. {\normalfont
  \bfseries 604} (2004)  88--107},
\href{http://arxiv.org/abs/astro-ph/0309465}{{\normalfont \ttfamily
  arXiv:astro-ph/0309465}}.

\bibitem{Newman:2012nw}
A.~B. Newman, T.~Treu, R.~S. Ellis, and D.~J. Sand, ``{\em {The Density
  Profiles of Massive, Relaxed Galaxy Clusters: II. Separating Luminous and
  Dark Matter in Cluster Cores}},''
  \href{http://dx.doi.org/10.1088/0004-637X/765/1/25}{Astrophys. J.
  {\normalfont \bfseries 765} (2013)  25},
\href{http://arxiv.org/abs/1209.1392}{{\normalfont \ttfamily arXiv:1209.1392}}.

\bibitem{BoylanKolchin:2011de}
M.~Boylan-Kolchin, J.~S. Bullock, and M.~Kaplinghat, ``{\em {Too big to fail?
  The puzzling darkness of massive Milky Way subhaloes}},''
  \href{http://dx.doi.org/10.1111/j.1745-3933.2011.01074.x}{Mon. Not. Roy.
  Astron. Soc. {\normalfont \bfseries 415} (2011)  L40},
\href{http://arxiv.org/abs/1103.0007}{{\normalfont \ttfamily arXiv:1103.0007}}.

\bibitem{Tollerud:2014zha}
E.~J. Tollerud, M.~Boylan-Kolchin, and J.~S. Bullock, ``{\em {M31 Satellite
  Masses Compared to LCDM Subhaloes}},''
  \href{http://dx.doi.org/10.1093/mnras/stu474}{Mon. Not. Roy. Astron. Soc.
  {\normalfont \bfseries 440} (2014) no.~4, 3511--3519},
\href{http://arxiv.org/abs/1403.6469}{{\normalfont \ttfamily arXiv:1403.6469}}.

\bibitem{Kirby:2014sya}
E.~N. Kirby, J.~S. Bullock, M.~Boylan-Kolchin, M.~Kaplinghat, and J.~G. Cohen,
  ``{\em {The dynamics of isolated Local Group galaxies}},''
  \href{http://dx.doi.org/10.1093/mnras/stu025}{Mon. Not. Roy. Astron. Soc.
  {\normalfont \bfseries 439} (2014) no.~1, 1015--1027},
\href{http://arxiv.org/abs/1401.1208}{{\normalfont \ttfamily arXiv:1401.1208}}.

\bibitem{Kaplinghat:2015aga}
M.~Kaplinghat, S.~Tulin, and H.-B. Yu, ``{\em {Dark Matter Halos as Particle
  Colliders: Unified Solution to Small-Scale Structure Puzzles from Dwarfs to
  Clusters}},'' \href{http://dx.doi.org/10.1103/PhysRevLett.116.041302}{Phys.
  Rev. Lett. {\normalfont \bfseries 116} (2016) no.~4, 041302},
\href{http://arxiv.org/abs/1508.03339}{{\normalfont \ttfamily
  arXiv:1508.03339}}.

\bibitem{Blok:2002tr}
W.~J. G.~d. Blok, A.~Bosma, and S.~S. McGaugh, ``{\em {Simulating observations
  of dark matter dominated galaxies: towards the optimal halo profile}},''
  \href{http://dx.doi.org/10.1046/j.1365-8711.2003.06330.x}{Mon. Not. Roy.
  Astron. Soc. {\normalfont \bfseries 340} (2003)  657--678},
\href{http://arxiv.org/abs/astro-ph/0212102}{{\normalfont \ttfamily
  arXiv:astro-ph/0212102}}.

\bibitem{Rhee:2003vw}
G.~Rhee, A.~Klypin, and O.~Valenzuela, ``{\em {The Rotation curves of dwarf
  galaxies: A Problem for cold dark matter?}},''
  \href{http://dx.doi.org/10.1086/425565}{Astrophys. J. {\normalfont \bfseries
  617} (2004)  1059--1076},
\href{http://arxiv.org/abs/astro-ph/0311020}{{\normalfont \ttfamily
  arXiv:astro-ph/0311020}}.

\bibitem{Gentile:2005de}
G.~Gentile, A.~Burkert, P.~Salucci, U.~Klein, and F.~Walter, ``{\em {The dwarf
  galaxy DDO 47 as a dark matter laboratory: testing cusps hiding in triaxial
  halos}},'' \href{http://dx.doi.org/10.1086/498939}{Astrophys. J. Lett.
  {\normalfont \bfseries 634} (2005)  L145--L148},
\href{http://arxiv.org/abs/astro-ph/0506538}{{\normalfont \ttfamily
  arXiv:astro-ph/0506538}}.

\bibitem{Spekkens:2005ik}
K.~Spekkens and R.~Giovanelli, ``{\em {The Cusp/core problem in Galactic halos:
  Long-slit spectra for a large dwarf galaxy sample}},''
  \href{http://dx.doi.org/10.1086/429592}{Astron. J. {\normalfont \bfseries
  129} (2005)  2119--2137},
\href{http://arxiv.org/abs/astro-ph/0502166}{{\normalfont \ttfamily
  arXiv:astro-ph/0502166}}.

\bibitem{Valenzuela:2005dh}
O.~Valenzuela, G.~Rhee, A.~Klypin, F.~Governato, G.~Stinson, T.~R. Quinn, and
  J.~Wadsley, ``{\em {Is there evidence for flat cores in the halos of dwarf
  galaxies?: the case of ngc 3109 and ngc 6822}},''
  \href{http://dx.doi.org/10.1086/508674}{Astrophys. J. {\normalfont \bfseries
  657} (2007)  773--789},
\href{http://arxiv.org/abs/astro-ph/0509644}{{\normalfont \ttfamily
  arXiv:astro-ph/0509644}}.

\bibitem{Dalcanton:2010bp}
J.~J. Dalcanton and A.~Stilp, ``{\em {Pressure Support in Galaxy Disks: Impact
  on Rotation Curves and Dark Matter Density Profiles}},''
  \href{http://dx.doi.org/10.1088/0004-637X/721/1/547}{Astrophys. J.
  {\normalfont \bfseries 721} (2010)  547--561},
\href{http://arxiv.org/abs/1007.2535}{{\normalfont \ttfamily arXiv:1007.2535}}.

\bibitem{Kormendy:2014ova}
J.~Kormendy and K.~C. Freeman, ``{\em {Scaling Laws for Dark Matter Halos in
  Late-type and Dwarf Spheroidal Galaxies}},''
  \href{http://dx.doi.org/10.3847/0004-637X/817/2/84}{Astrophys. J.
  {\normalfont \bfseries 817} (2016) no.~2, 84},
\href{http://arxiv.org/abs/1411.2170}{{\normalfont \ttfamily arXiv:1411.2170}}.

\bibitem{2016MNRAS.462.3628R}
J.~I. {Read}, G.~{Iorio}, O.~{Agertz}, and F.~{Fraternali},
  \href{http://dx.doi.org/10.1093/mnras/stw1876}{``{\em {Understanding the
  shape and diversity of dwarf galaxy rotation curves in
  {$\Lambda$}CDM}},''Mon. Not. Roy. Astron. Soc. {\normalfont \bfseries 462}
  (Nov., 2016)  3628--3645},
  \href{http://arxiv.org/abs/1601.05821}{{\normalfont \ttfamily
  arXiv:1601.05821}}.

\bibitem{Maccio:2016egb}
A.~V. Macci\`{o}, S.~M. Udrescu, A.~A. Dutton, A.~Obreja, L.~Wang, G.~R.
  Stinson, and X.~Kang, ``{\em {NIHAO X: reconciling the local galaxy velocity
  function with cold dark matter via mock HI observations}},''
  \href{http://dx.doi.org/10.1093/mnrasl/slw147}{Mon. Not. Roy. Astron. Soc.
  {\normalfont \bfseries 463} (2016) no.~1, L69--L73},
\href{http://arxiv.org/abs/1607.01028}{{\normalfont \ttfamily
  arXiv:1607.01028}}.

\bibitem{2017A&A...601A...1P}
E.~{Papastergis} and A.~A. {Ponomareva},
  \href{http://dx.doi.org/10.1051/0004-6361/201629546}{``{\em {Testing core
  creation in hydrodynamical simulations using the HI kinematics of field
  dwarfs}},''Astronomy \& Astrophysics {\normalfont \bfseries 601} (May, 2017)
  A1}, \href{http://arxiv.org/abs/1608.05214}{{\normalfont \ttfamily
  arXiv:1608.05214}}.

\bibitem{Brooks:2017rfe}
A.~M. Brooks, E.~Papastergis, C.~R. Christensen, F.~Governato, A.~Stilp, T.~R.
  Quinn, and J.~Wadsley, ``{\em {How to Reconcile the Observed Velocity
  Function of Galaxies with Theory}},''
  \href{http://dx.doi.org/10.3847/1538-4357/aa9576}{Astrophys. J. {\normalfont
  \bfseries 850} (2017) no.~1, 97},
\href{http://arxiv.org/abs/1701.07835}{{\normalfont \ttfamily
  arXiv:1701.07835}}.

\bibitem{Oman:2017vkl}
K.~A. Oman, A.~Marasco, J.~F. Navarro, C.~S. Frenk, J.~Schaye, and
  A.~Benítez-Llambay, ``{\em {Apparent cores and non-circular motions in the
  HI discs of simulated galaxies}},''
\href{http://arxiv.org/abs/1706.07478}{{\normalfont \ttfamily
  arXiv:1706.07478}}.

\bibitem{2018MNRAS.474.1398G}
A.~{Genina}, A.~{Ben{\'{\i}}tez-Llambay}, C.~S. {Frenk}, S.~{Cole},
  A.~{Fattahi}, J.~F. {Navarro}, K.~A. {Oman}, T.~{Sawala}, and T.~{Theuns},
  \href{http://dx.doi.org/10.1093/mnras/stx2855}{``{\em {The core-cusp problem:
  a matter of perspective}},''"Mon. Not. Roy. Astron. Soc." {\normalfont
  \bfseries 474} (Feb., 2018)  1398--1411}.

\bibitem{Read:2018pft}
J.~I. Read, M.~G. Walker, and P.~Steger, ``{\em {The case for a cold dark
  matter cusp in Draco}},''
\href{http://arxiv.org/abs/1805.06934}{{\normalfont \ttfamily
  arXiv:1805.06934}}.

\bibitem{Navarro:1996bv}
J.~F. Navarro, V.~R. Eke, and C.~S. Frenk, ``{\em {The cores of dwarf galaxy
  halos}},'' \href{http://dx.doi.org/10.1093/mnras/283.3.72L,
  10.1093/mnras/283.3.L72}{Mon. Not. Roy. Astron. Soc. {\normalfont \bfseries
  283} (1996)  L72--L78},
\href{http://arxiv.org/abs/astro-ph/9610187}{{\normalfont \ttfamily
  arXiv:astro-ph/9610187}}.

\bibitem{Gelato:1998hb}
S.~Gelato and J.~Sommer-Larsen, ``{\em {On ddo154 and cold dark matter halo
  profiles}},'' \href{http://dx.doi.org/10.1046/j.1365-8711.1999.02223.x}{Mon.
  Not. Roy. Astron. Soc. {\normalfont \bfseries 303} (1999)  321--328},
\href{http://arxiv.org/abs/astro-ph/9806289}{{\normalfont \ttfamily
  arXiv:astro-ph/9806289}}.

\bibitem{Binney:2000zt}
J.~Binney, O.~Gerhard, and J.~Silk, ``{\em {The Dark matter problem in disk
  galaxies}},'' \href{http://dx.doi.org/10.1046/j.1365-8711.2001.04024.x}{Mon.
  Not. Roy. Astron. Soc. {\normalfont \bfseries 321} (2001)  471},
\href{http://arxiv.org/abs/astro-ph/0003199}{{\normalfont \ttfamily
  arXiv:astro-ph/0003199}}.

\bibitem{Gnedin:2001ec}
O.~Y. Gnedin and H.~Zhao, ``{\em {Maximum feedback and dark matter profiles of
  dwarf galaxies}},''
  \href{http://dx.doi.org/10.1046/j.1365-8711.2002.05361.x}{Mon. Not. Roy.
  Astron. Soc. {\normalfont \bfseries 333} (2002)  299},
\href{http://arxiv.org/abs/astro-ph/0108108}{{\normalfont \ttfamily
  arXiv:astro-ph/0108108}}.

\bibitem{Read:2018fxs}
J.~I. Read, M.~G. Walker, and P.~Steger, ``{\em {Dark matter heats up in dwarf
  galaxies}},''
\href{http://arxiv.org/abs/1808.06634}{{\normalfont \ttfamily
  arXiv:1808.06634}}.

\bibitem{ElZant:2001re}
A.~El-Zant, I.~Shlosman, and Y.~Hoffman, ``{\em {Dark halos: the flattening of
  the density cusp by dynamical friction}},''
  \href{http://dx.doi.org/10.1086/322516}{Astrophys. J. {\normalfont \bfseries
  560} (2001)  636},
\href{http://arxiv.org/abs/astro-ph/0103386}{{\normalfont \ttfamily
  arXiv:astro-ph/0103386}}.

\bibitem{Weinberg:2001gm}
M.~D. Weinberg and N.~Katz, ``{\em {Bar-driven dark halo evolution: a
  resolution of the cusp-core controversy}},''
  \href{http://dx.doi.org/10.1086/343847}{Astrophys. J. {\normalfont \bfseries
  580} (2002)  627--633},
\href{http://arxiv.org/abs/astro-ph/0110632}{{\normalfont \ttfamily
  arXiv:astro-ph/0110632}}.

\bibitem{Ahn:2004xt}
K.-J. Ahn and P.~R. Shapiro, ``{\em {Formation and evolution of the
  self-interacting dark matter halos}},''
  \href{http://dx.doi.org/10.1111/j.1365-2966.2005.09492.x}{Mon. Not. Roy.
  Astron. Soc. {\normalfont \bfseries 363} (2005)  1092--1124},
\href{http://arxiv.org/abs/astro-ph/0412169}{{\normalfont \ttfamily
  arXiv:astro-ph/0412169}}.

\bibitem{Tonini:2006gwz}
C.~Tonini and A.~Lapi, ``{\em {Angular momentum transfer in dark matter halos:
  erasing the cusp}},'' \href{http://dx.doi.org/10.1086/506431}{Astrophys. J.
  {\normalfont \bfseries 649} (2006)  591--598},
\href{http://arxiv.org/abs/astro-ph/0603051}{{\normalfont \ttfamily
  arXiv:astro-ph/0603051}}.

\bibitem{Martizzi:2011aa}
D.~Martizzi, R.~Teyssier, B.~Moore, and T.~Wentz, ``{\em {The effects of baryon
  physics, black holes and AGN feedback on the mass distribution in clusters of
  galaxies}},'' \href{http://dx.doi.org/10.1111/j.1365-2966.2012.20879.x}{Mon.
  Not. Roy. Astron. Soc. {\normalfont \bfseries 422} (2012)  3081},
\href{http://arxiv.org/abs/1112.2752}{{\normalfont \ttfamily arXiv:1112.2752}}.

\bibitem{Spergel:1999mh}
D.~N. Spergel and P.~J. Steinhardt, ``{\em {Observational evidence for
  selfinteracting cold dark matter}},''
  \href{http://dx.doi.org/10.1103/PhysRevLett.84.3760}{Phys. Rev. Lett.
  {\normalfont \bfseries 84} (2000)  3760--3763},
\href{http://arxiv.org/abs/astro-ph/9909386}{{\normalfont \ttfamily
  arXiv:astro-ph/9909386}}.

\bibitem{Dave:2000ar}
R.~Dave, D.~N. Spergel, P.~J. Steinhardt, and B.~D. Wandelt, ``{\em {Halo
  properties in cosmological simulations of selfinteracting cold dark
  matter}},'' \href{http://dx.doi.org/10.1086/318417}{Astrophys. J.
  {\normalfont \bfseries 547} (2001)  574--589},
\href{http://arxiv.org/abs/astro-ph/0006218}{{\normalfont \ttfamily
  arXiv:astro-ph/0006218}}.

\bibitem{Vogelsberger:2012ku}
M.~Vogelsberger, J.~Zavala, and A.~Loeb, ``{\em {Subhaloes in Self-Interacting
  Galactic Dark Matter Haloes}},''
  \href{http://dx.doi.org/10.1111/j.1365-2966.2012.21182.x}{Mon. Not. Roy.
  Astron. Soc. {\normalfont \bfseries 423} (2012)  3740},
\href{http://arxiv.org/abs/1201.5892}{{\normalfont \ttfamily arXiv:1201.5892}}.

\bibitem{Rocha:2012jg}
M.~Rocha, A.~H.~G. Peter, J.~S. Bullock, M.~Kaplinghat, S.~Garrison-Kimmel,
  J.~Onorbe, and L.~A. Moustakas, ``{\em {Cosmological Simulations with
  Self-Interacting Dark Matter I: Constant Density Cores and Substructure}},''
  \href{http://dx.doi.org/10.1093/mnras/sts514}{Mon. Not. Roy. Astron. Soc.
  {\normalfont \bfseries 430} (2013)  81--104},
\href{http://arxiv.org/abs/1208.3025}{{\normalfont \ttfamily arXiv:1208.3025}}.

\bibitem{Peter:2012jh}
A.~H.~G. Peter, M.~Rocha, J.~S. Bullock, and M.~Kaplinghat, ``{\em
  {Cosmological Simulations with Self-Interacting Dark Matter II: Halo Shapes
  vs. Observations}},'' \href{http://dx.doi.org/10.1093/mnras/sts535}{Mon. Not.
  Roy. Astron. Soc. {\normalfont \bfseries 430} (2013)  105},
\href{http://arxiv.org/abs/1208.3026}{{\normalfont \ttfamily arXiv:1208.3026}}.

\bibitem{Elbert:2014bma}
O.~D. Elbert, J.~S. Bullock, S.~Garrison-Kimmel, M.~Rocha, J.~Oñorbe, and
  A.~H.~G. Peter, ``{\em {Core formation in dwarf haloes with self-interacting
  dark matter: no fine-tuning necessary}},''
  \href{http://dx.doi.org/10.1093/mnras/stv1470}{Mon. Not. Roy. Astron. Soc.
  {\normalfont \bfseries 453} (2015) no.~1, 29--37},
\href{http://arxiv.org/abs/1412.1477}{{\normalfont \ttfamily arXiv:1412.1477}}.

\bibitem{Fry:2015rta}
A.~B. Fry, F.~Governato, A.~Pontzen, T.~Quinn, M.~Tremmel, L.~Anderson,
  H.~Menon, A.~M. Brooks, and J.~Wadsley, ``{\em {All about baryons: revisiting
  SIDM predictions at small halo masses}},''
  \href{http://dx.doi.org/10.1093/mnras/stv1330}{Mon. Not. Roy. Astron. Soc.
  {\normalfont \bfseries 452} (2015) no.~2, 1468--1479},
\href{http://arxiv.org/abs/1501.00497}{{\normalfont \ttfamily
  arXiv:1501.00497}}.

\bibitem{Tulin:2017ara}
S.~Tulin and H.-B. Yu, ``{\em {Dark Matter Self-interactions and Small Scale
  Structure}},'' \href{http://dx.doi.org/10.1016/j.physrep.2017.11.004}{Phys.
  Rept. {\normalfont \bfseries 730} (2018)  1--57},
\href{http://arxiv.org/abs/1705.02358}{{\normalfont \ttfamily
  arXiv:1705.02358}}.

\bibitem{Feng:2009hw}
J.~L. Feng, M.~Kaplinghat, and H.-B. Yu, ``{\em {Halo Shape and Relic Density
  Exclusions of Sommerfeld-Enhanced Dark Matter Explanations of Cosmic Ray
  Excesses}},'' \href{http://dx.doi.org/10.1103/PhysRevLett.104.151301}{Phys.
  Rev. Lett. {\normalfont \bfseries 104} (2010)  151301},
\href{http://arxiv.org/abs/0911.0422}{{\normalfont \ttfamily arXiv:0911.0422}}.

\bibitem{McDermott:2017vyk}
S.~D. McDermott, ``{\em {Is Self-Interacting Dark Matter Undergoing Dark
  Fusion?}},'' \href{http://dx.doi.org/10.1103/PhysRevLett.120.221806}{Phys.
  Rev. Lett. {\normalfont \bfseries 120} (2018) no.~22, 221806},
\href{http://arxiv.org/abs/1711.00857}{{\normalfont \ttfamily
  arXiv:1711.00857}}.

\bibitem{Vogelsberger:2018bok}
M.~Vogelsberger, J.~Zavala, K.~Schutz, and T.~R. Slatyer, ``{\em {Evaporating
  the Milky Way halo and its satellites with inelastic self-interacting dark
  matter}},''
\href{http://arxiv.org/abs/1805.03203}{{\normalfont \ttfamily
  arXiv:1805.03203}}.

\bibitem{Kamada:2017gfc}
A.~Kamada, H.~J. Kim, H.~Kim, and T.~Sekiguchi, ``{\em {Self-Heating Dark
  Matter via Semiannihilation}},''
  \href{http://dx.doi.org/10.1103/PhysRevLett.120.131802}{Phys. Rev. Lett.
  {\normalfont \bfseries 120} (2018) no.~13, 131802},
\href{http://arxiv.org/abs/1707.09238}{{\normalfont \ttfamily
  arXiv:1707.09238}}.

\bibitem{Chu:2018nki}
X.~Chu and C.~Garcia-Cely, ``{\em {Core formation from self-heating dark
  matter}},'' \href{http://dx.doi.org/10.1088/1475-7516/2018/07/013}{JCAP
  {\normalfont \bfseries 1807} (2018) no.~07, 013},
\href{http://arxiv.org/abs/1803.09762}{{\normalfont \ttfamily
  arXiv:1803.09762}}.

\bibitem{Kamada:2018hte}
A.~Kamada, H.~J. Kim, and H.~Kim, ``{\em {Self-heating of Strongly Interacting
  Massive Particles}},''
  \href{http://dx.doi.org/10.1103/PhysRevD.98.023509}{Phys. Rev. {\normalfont
  \bfseries D98} (2018) no.~2, 023509},
\href{http://arxiv.org/abs/1805.05648}{{\normalfont \ttfamily
  arXiv:1805.05648}}.

\bibitem{Griest:1990kh}
K.~Griest and D.~Seckel, ``{\em {Three exceptions in the calculation of relic
  abundances}},''
\href{http://dx.doi.org/10.1103/PhysRevD.43.3191}{Phys. Rev. {\normalfont
  \bfseries D43} (1991)  3191--3203}.

\bibitem{Gondolo:1990dk}
P.~Gondolo and G.~Gelmini, ``{\em {Cosmic abundances of stable particles:
  Improved analysis}},''
\href{http://dx.doi.org/10.1016/0550-3213(91)90438-4}{Nucl. Phys. {\normalfont
  \bfseries B360} (1991)  145--179}.

\bibitem{Jungman:1995df}
G.~Jungman, M.~Kamionkowski, and K.~Griest, ``{\em {Supersymmetric dark
  matter}},'' \href{http://dx.doi.org/10.1016/0370-1573(95)00058-5}{Phys. Rept.
  {\normalfont \bfseries 267} (1996)  195--373},
\href{http://arxiv.org/abs/hep-ph/9506380}{{\normalfont \ttfamily
  arXiv:hep-ph/9506380}}.

\bibitem{Feldman:2008xs}
D.~Feldman, Z.~Liu, and P.~Nath, ``{\em {PAMELA Positron Excess as a Signal
  from the Hidden Sector}},''
  \href{http://dx.doi.org/10.1103/PhysRevD.79.063509}{Phys. Rev. {\normalfont
  \bfseries D79} (2009)  063509},
\href{http://arxiv.org/abs/0810.5762}{{\normalfont \ttfamily arXiv:0810.5762}}.

\bibitem{Pospelov:2008jd}
M.~Pospelov and A.~Ritz, ``{\em {Astrophysical Signatures of Secluded Dark
  Matter}},'' \href{http://dx.doi.org/10.1016/j.physletb.2008.12.012}{Phys.
  Lett. {\normalfont \bfseries B671} (2009)  391--397},
\href{http://arxiv.org/abs/0810.1502}{{\normalfont \ttfamily arXiv:0810.1502}}.

\bibitem{Ibe:2008ye}
M.~Ibe, H.~Murayama, and T.~T. Yanagida, ``{\em {Breit-Wigner Enhancement of
  Dark Matter Annihilation}},''
  \href{http://dx.doi.org/10.1103/PhysRevD.79.095009}{Phys. Rev. {\normalfont
  \bfseries D79} (2009)  095009},
\href{http://arxiv.org/abs/0812.0072}{{\normalfont \ttfamily arXiv:0812.0072}}.

\bibitem{MarchRussell:2008tu}
J.~D. March-Russell and S.~M. West, ``{\em {WIMPonium and Boost Factors for
  Indirect Dark Matter Detection}},''
  \href{http://dx.doi.org/10.1016/j.physletb.2009.04.010}{Phys. Lett.
  {\normalfont \bfseries B676} (2009)  133--139},
\href{http://arxiv.org/abs/0812.0559}{{\normalfont \ttfamily arXiv:0812.0559}}.

\bibitem{Guo:2009aj}
W.-L. Guo and Y.-L. Wu, ``{\em {Enhancement of Dark Matter Annihilation via
  Breit-Wigner Resonance}},''
  \href{http://dx.doi.org/10.1103/PhysRevD.79.055012}{Phys. Rev. {\normalfont
  \bfseries D79} (2009)  055012},
\href{http://arxiv.org/abs/0901.1450}{{\normalfont \ttfamily arXiv:0901.1450}}.

\bibitem{Ibe:2009dx}
M.~Ibe, Y.~Nakayama, H.~Murayama, and T.~T. Yanagida, ``{\em {Nambu-Goldstone
  Dark Matter and Cosmic Ray Electron and Positron Excess}},''
  \href{http://dx.doi.org/10.1088/1126-6708/2009/04/087}{JHEP {\normalfont
  \bfseries 04} (2009)  087},
\href{http://arxiv.org/abs/0902.2914}{{\normalfont \ttfamily arXiv:0902.2914}}.

\bibitem{Kakizaki:2005en}
M.~Kakizaki, S.~Matsumoto, Y.~Sato, and M.~Senami, ``{\em {Significant effects
  of second KK particles on LKP dark matter physics}},''
  \href{http://dx.doi.org/10.1103/PhysRevD.71.123522}{Phys. Rev. {\normalfont
  \bfseries D71} (2005)  123522},
\href{http://arxiv.org/abs/hep-ph/0502059}{{\normalfont \ttfamily
  arXiv:hep-ph/0502059}}.

\bibitem{Arina:2014fna}
C.~Arina, T.~Bringmann, J.~Silk, and M.~Vollmann, ``{\em {Enhanced Line Signals
  from Annihilating Kaluza-Klein Dark Matter}},''
  \href{http://dx.doi.org/10.1103/PhysRevD.90.083506}{Phys. Rev. {\normalfont
  \bfseries D90} (2014) no.~8, 083506},
\href{http://arxiv.org/abs/1409.0007}{{\normalfont \ttfamily arXiv:1409.0007}}.

\bibitem{Ibe:2009mk}
M.~Ibe and H.-b. Yu, ``{\em {Distinguishing Dark Matter Annihilation
  Enhancement Scenarios via Halo Shapes}},''
  \href{http://dx.doi.org/10.1016/j.physletb.2010.07.026}{Phys. Lett.
  {\normalfont \bfseries B692} (2010)  70--73},
\href{http://arxiv.org/abs/0912.5425}{{\normalfont \ttfamily arXiv:0912.5425}}.

\bibitem{Braaten:2013tza}
E.~Braaten and H.~W. Hammer, ``{\em {Universal Two-body Physics in Dark Matter
  near an S-wave Resonance}},''
  \href{http://dx.doi.org/10.1103/PhysRevD.88.063511}{Phys. Rev. {\normalfont
  \bfseries D88} (2013)  063511},
\href{http://arxiv.org/abs/1303.4682}{{\normalfont \ttfamily arXiv:1303.4682}}.

\bibitem{Duch:2017nbe}
M.~Duch and B.~Grzadkowski, ``{\em {Resonance enhancement of dark matter
  interactions: the case for early kinetic decoupling and velocity dependent
  resonance width}},'' \href{http://dx.doi.org/10.1007/JHEP09(2017)159}{JHEP
  {\normalfont \bfseries 09} (2017)  159},
\href{http://arxiv.org/abs/1705.10777}{{\normalfont \ttfamily
  arXiv:1705.10777}}.

\bibitem{Braaten:2018xuw}
E.~Braaten, D.~Kang, and R.~Laha, ``{\em {Production of dark-matter bound
  states in the early universe by three-body recombination}},''
\href{http://arxiv.org/abs/1806.00609}{{\normalfont \ttfamily
  arXiv:1806.00609}}.

\bibitem{KuziodeNaray:2007qi}
R.~Kuzio~de Naray, S.~S. McGaugh, and W.~J.~G. de~Blok, ``{\em {Mass Models for
  Low Surface Brightness Galaxies with High Resolution Optical Velocity
  Fields}},'' \href{http://dx.doi.org/10.1086/527543}{Astrophys. J.
  {\normalfont \bfseries 676} (2008)  920--943},
\href{http://arxiv.org/abs/0712.0860}{{\normalfont \ttfamily arXiv:0712.0860}}.

\bibitem{Oh:2010ea}
S.-H. Oh, W.~J.~G. de~Blok, E.~Brinks, F.~Walter, and R.~C. Kennicutt, Jr,
  ``{\em {Dark and luminous matter in THINGS dwarf galaxies}},''
  \href{http://dx.doi.org/10.1088/0004-6256/141/6/193}{Astron. J. {\normalfont
  \bfseries 141} (2011)  193},
\href{http://arxiv.org/abs/1011.0899}{{\normalfont \ttfamily arXiv:1011.0899}}.

\bibitem{Elbert:2016dbb}
O.~D. Elbert, J.~S. Bullock, M.~Kaplinghat, S.~Garrison-Kimmel, A.~S. Graus,
  and M.~Rocha, ``{\em {A Testable Conspiracy: Simulating Baryonic Effects on
  Self-Interacting Dark Matter Halos}},''
  \href{http://dx.doi.org/10.3847/1538-4357/aa9710}{Astrophys. J. {\normalfont
  \bfseries 853} (2018) no.~2, 109},
\href{http://arxiv.org/abs/1609.08626}{{\normalfont \ttfamily
  arXiv:1609.08626}}.

\bibitem{Valli:2017ktb}
M.~Valli and H.-B. Yu, ``{\em {Dark matter self-interactions from the internal
  dynamics of dwarf spheroidals}},''
\href{http://arxiv.org/abs/1711.03502}{{\normalfont \ttfamily
  arXiv:1711.03502}}.

\bibitem{sokolenko:2018noz}
A.~Sokolenko, K.~Bondarenko, T.~Brinckmann, J.~Zavala, M.~Vogelsberger,
  T.~Bringmann, and A.~Boyarsky, ``{\em {Towards an improved model of
  self-interacting dark matter haloes}},''
\href{http://arxiv.org/abs/1806.11539}{{\normalfont \ttfamily
  arXiv:1806.11539}}.

\bibitem{Randall:2007ph}
S.~W. Randall, M.~Markevitch, D.~Clowe, A.~H. Gonzalez, and M.~Bradac, ``{\em
  {Constraints on the Self-Interaction Cross-Section of Dark Matter from
  Numerical Simulations of the Merging Galaxy Cluster 1E 0657-56}},''
  \href{http://dx.doi.org/10.1086/587859}{Astrophys. J. {\normalfont \bfseries
  679} (2008)  1173--1180},
\href{http://arxiv.org/abs/0704.0261}{{\normalfont \ttfamily arXiv:0704.0261}}.

\bibitem{Robertson:2016xjh}
A.~Robertson, R.~Massey, and V.~Eke, ``{\em {What does the Bullet Cluster tell
  us about self-interacting dark matter?}},''
  \href{http://dx.doi.org/10.1093/mnras/stw2670}{Mon. Not. Roy. Astron. Soc.
  {\normalfont \bfseries 465} (2017) no.~1, 569--587},
\href{http://arxiv.org/abs/1605.04307}{{\normalfont \ttfamily
  arXiv:1605.04307}}.

\bibitem{Breit:1936zzb}
G.~Breit and E.~Wigner, ``{\em {Capture of Slow Neutrons}},''
\href{http://dx.doi.org/10.1103/PhysRev.49.519}{Phys. Rev. {\normalfont
  \bfseries 49} (1936)  519--531}.

\bibitem{Kahlhoefer:2017umn}
F.~Kahlhoefer, K.~Schmidt-Hoberg, and S.~Wild, ``{\em {Dark matter
  self-interactions from a general spin-0 mediator}},''
  \href{http://dx.doi.org/10.1088/1475-7516/2017/08/003}{JCAP {\normalfont
  \bfseries 1708} (2017) no.~08, 003},
\href{http://arxiv.org/abs/1704.02149}{{\normalfont \ttfamily
  arXiv:1704.02149}}.

\bibitem{Francis:2018xjd}
A.~Francis, R.~J. Hudspith, R.~Lewis, and S.~Tulin, ``{\em {Dark Matter from
  Strong Dynamics: The Minimal Theory of Dark Baryons}},''
\href{http://arxiv.org/abs/1809.09117}{{\normalfont \ttfamily
  arXiv:1809.09117}}.

\bibitem{Dolgov:1980uu}
A.~D. Dolgov, ``{\em {ON CONCENTRATION OF RELICT THETA PARTICLES. (IN
  RUSSIAN)}},''
Yad. Fiz. {\normalfont \bfseries 31} (1980)  1522--1528.

\bibitem{Carlson:1992fn}
E.~D. Carlson, M.~E. Machacek, and L.~J. Hall, ``{\em {Self-interacting dark
  matter}},''
\href{http://dx.doi.org/10.1086/171833}{Astrophys. J. {\normalfont \bfseries
  398} (1992)  43--52}.

\bibitem{Hochberg:2014dra}
Y.~Hochberg, E.~Kuflik, T.~Volansky, and J.~G. Wacker, ``{\em {Mechanism for
  Thermal Relic Dark Matter of Strongly Interacting Massive Particles}},''
  \href{http://dx.doi.org/10.1103/PhysRevLett.113.171301}{Phys. Rev. Lett.
  {\normalfont \bfseries 113} (2014)  171301},
\href{http://arxiv.org/abs/1402.5143}{{\normalfont \ttfamily arXiv:1402.5143}}.

\bibitem{Yamanaka:2014pva}
N.~Yamanaka, S.~Fujibayashi, S.~Gongyo, and H.~Iida, ``{\em {Dark matter in the
  hidden gauge theory}},''
\href{http://arxiv.org/abs/1411.2172}{{\normalfont \ttfamily arXiv:1411.2172}}.

\bibitem{Hochberg:2014kqa}
Y.~Hochberg, E.~Kuflik, H.~Murayama, T.~Volansky, and J.~G. Wacker, ``{\em
  {Model for Thermal Relic Dark Matter of Strongly Interacting Massive
  Particles}},'' \href{http://dx.doi.org/10.1103/PhysRevLett.115.021301}{Phys.
  Rev. Lett. {\normalfont \bfseries 115} (2015) no.~2, 021301},
\href{http://arxiv.org/abs/1411.3727}{{\normalfont \ttfamily arXiv:1411.3727}}.

\bibitem{Bernal:2015bla}
N.~Bernal, C.~Garcia-Cely, and R.~Rosenfeld, ``{\em {WIMP and SIMP Dark Matter
  from the Spontaneous Breaking of a Global Group}},''
  \href{http://dx.doi.org/10.1088/1475-7516/2015/04/012}{JCAP {\normalfont
  \bfseries 1504} (2015) no.~04, 012},
\href{http://arxiv.org/abs/1501.01973}{{\normalfont \ttfamily
  arXiv:1501.01973}}.

\bibitem{Bernal:2015lbl}
N.~Bernal, C.~Garcia-Cely, and R.~Rosenfeld, ``{\em {$\mathbb{Z}_3$ WIMP and
  SIMP Dark Matter from a Global U(1) Breaking}},''
\href{http://dx.doi.org/10.1016/j.nuclphysbps.2015.11.001}{Nucl. Part. Phys.
  Proc. {\normalfont \bfseries 267-269} (2015)  353--355}.

\bibitem{Lee:2015gsa}
H.~M. Lee and M.-S. Seo, ``{\em {Communication with SIMP dark mesons via Z'
  -portal}},'' \href{http://dx.doi.org/10.1016/j.physletb.2015.07.013}{Phys.
  Lett. {\normalfont \bfseries B748} (2015)  316--322},
\href{http://arxiv.org/abs/1504.00745}{{\normalfont \ttfamily
  arXiv:1504.00745}}.

\bibitem{Choi:2015bya}
S.-M. Choi and H.~M. Lee, ``{\em {SIMP dark matter with gauged $\mathbb{Z}_{3}$
  symmetry}},'' \href{http://dx.doi.org/10.1007/JHEP09(2015)063}{JHEP
  {\normalfont \bfseries 09} (2015)  063},
\href{http://arxiv.org/abs/1505.00960}{{\normalfont \ttfamily
  arXiv:1505.00960}}.

\bibitem{Hansen:2015yaa}
M.~Hansen, K.~Langæble, and F.~Sannino, ``{\em {SIMP model at NNLO in chiral
  perturbation theory}},''
  \href{http://dx.doi.org/10.1103/PhysRevD.92.075036}{Phys. Rev. {\normalfont
  \bfseries D92} (2015) no.~7, 075036},
\href{http://arxiv.org/abs/1507.01590}{{\normalfont \ttfamily
  arXiv:1507.01590}}.

\bibitem{Bernal:2015xba}
N.~Bernal and X.~Chu, ``{\em {$\mathbb{Z}_2$ SIMP Dark Matter}},''
  \href{http://dx.doi.org/10.1088/1475-7516/2016/01/006}{JCAP {\normalfont
  \bfseries 1601} (2016)  006},
\href{http://arxiv.org/abs/1510.08527}{{\normalfont \ttfamily
  arXiv:1510.08527}}.

\bibitem{Kuflik:2015isi}
E.~Kuflik, M.~Perelstein, N.~R.-L. Lorier, and Y.-D. Tsai, ``{\em {Elastically
  Decoupling Dark Matter}},''
  \href{http://dx.doi.org/10.1103/PhysRevLett.116.221302}{Phys. Rev. Lett.
  {\normalfont \bfseries 116} (2016) no.~22, 221302},
\href{http://arxiv.org/abs/1512.04545}{{\normalfont \ttfamily
  arXiv:1512.04545}}.

\bibitem{Hochberg:2015vrg}
Y.~Hochberg, E.~Kuflik, and H.~Murayama, ``{\em {SIMP Spectroscopy}},''
  \href{http://dx.doi.org/10.1007/JHEP05(2016)090}{JHEP {\normalfont \bfseries
  05} (2016)  090},
\href{http://arxiv.org/abs/1512.07917}{{\normalfont \ttfamily
  arXiv:1512.07917}}.

\bibitem{Choi:2016hid}
S.-M. Choi and H.~M. Lee, ``{\em {Resonant SIMP dark matter}},''
  \href{http://dx.doi.org/10.1016/j.physletb.2016.04.055}{Phys. Lett.
  {\normalfont \bfseries B758} (2016)  47--53},
\href{http://arxiv.org/abs/1601.03566}{{\normalfont \ttfamily
  arXiv:1601.03566}}.

\bibitem{Pappadopulo:2016pkp}
D.~Pappadopulo, J.~T. Ruderman, and G.~Trevisan, ``{\em {Dark matter freeze-out
  in a nonrelativistic sector}},''
  \href{http://dx.doi.org/10.1103/PhysRevD.94.035005}{Phys. Rev. {\normalfont
  \bfseries D94} (2016) no.~3, 035005},
\href{http://arxiv.org/abs/1602.04219}{{\normalfont \ttfamily
  arXiv:1602.04219}}.

\bibitem{Farina:2016llk}
M.~Farina, D.~Pappadopulo, J.~T. Ruderman, and G.~Trevisan, ``{\em {Phases of
  Cannibal Dark Matter}},''
  \href{http://dx.doi.org/10.1007/JHEP12(2016)039}{JHEP {\normalfont \bfseries
  12} (2016)  039},
\href{http://arxiv.org/abs/1607.03108}{{\normalfont \ttfamily
  arXiv:1607.03108}}.

\bibitem{Choi:2016tkj}
S.-M. Choi, Y.-J. Kang, and H.~M. Lee, ``{\em {On thermal production of
  self-interacting dark matter}},''
  \href{http://dx.doi.org/10.1007/JHEP12(2016)099}{JHEP {\normalfont \bfseries
  12} (2016)  099},
\href{http://arxiv.org/abs/1610.04748}{{\normalfont \ttfamily
  arXiv:1610.04748}}.

\bibitem{Dey:2016qgf}
U.~K. Dey, T.~N. Maity, and T.~S. Ray, ``{\em {Light Dark Matter through
  Assisted Annihilation}},''
\href{http://arxiv.org/abs/1612.09074}{{\normalfont \ttfamily
  arXiv:1612.09074}}.

\bibitem{Cline:2017tka}
J.~Cline, H.~Liu, T.~Slatyer, and W.~Xue, ``{\em {Enabling Forbidden Dark
  Matter}},''
\href{http://arxiv.org/abs/1702.07716}{{\normalfont \ttfamily
  arXiv:1702.07716}}.

\bibitem{Choi:2017mkk}
S.-M. Choi, H.~M. Lee, and M.-S. Seo, ``{\em {Cosmic abundances of SIMP dark
  matter}},'' \href{http://dx.doi.org/10.1007/JHEP04(2017)154}{JHEP
  {\normalfont \bfseries 04} (2017)  154},
\href{http://arxiv.org/abs/1702.07860}{{\normalfont \ttfamily
  arXiv:1702.07860}}.

\bibitem{Choi:2017zww}
S.-M. Choi, Y.~Hochberg, E.~Kuflik, H.~M. Lee, Y.~Mambrini, H.~Murayama, and
  M.~Pierre, ``{\em {Vector SIMP dark matter}},''
\href{http://arxiv.org/abs/1707.01434}{{\normalfont \ttfamily
  arXiv:1707.01434}}.

\bibitem{Chu:2017msm}
X.~Chu and C.~Garcia-Cely, ``{\em {Self-interacting Spin-2 Dark Matter}},''
  \href{http://dx.doi.org/10.1103/PhysRevD.96.103519}{Phys. Rev. {\normalfont
  \bfseries D96} (2017) no.~10, 103519},
\href{http://arxiv.org/abs/1708.06764}{{\normalfont \ttfamily
  arXiv:1708.06764}}.

\bibitem{Choi:2018iit}
S.-M. Choi, H.~M. Lee, P.~Ko, and A.~Natale, ``{\em {Resolving phenomenological
  problems with strongly-interacting-massive-particle models with dark vector
  resonances}},'' \href{http://dx.doi.org/10.1103/PhysRevD.98.015034}{Phys.
  Rev. {\normalfont \bfseries D98} (2018) no.~1, 015034},
\href{http://arxiv.org/abs/1801.07726}{{\normalfont \ttfamily
  arXiv:1801.07726}}.

\bibitem{McDonald:2001vt}
J.~McDonald, ``{\em {Thermally generated gauge singlet scalars as
  selfinteracting dark matter}},''
  \href{http://dx.doi.org/10.1103/PhysRevLett.88.091304}{Phys. Rev. Lett.
  {\normalfont \bfseries 88} (2002)  091304},
\href{http://arxiv.org/abs/hep-ph/0106249}{{\normalfont \ttfamily
  arXiv:hep-ph/0106249}}.

\bibitem{Hall:2009bx}
L.~J. Hall, K.~Jedamzik, J.~March-Russell, and S.~M. West, ``{\em {Freeze-In
  Production of FIMP Dark Matter}},''
  \href{http://dx.doi.org/10.1007/JHEP03(2010)080}{JHEP {\normalfont \bfseries
  03} (2010)  080},
\href{http://arxiv.org/abs/0911.1120}{{\normalfont \ttfamily arXiv:0911.1120}}.

\bibitem{Bernal:2015ova}
N.~Bernal, X.~Chu, C.~Garcia-Cely, T.~Hambye, and B.~Zaldivar, ``{\em
  {Production Regimes for Self-Interacting Dark Matter}},''
  \href{http://dx.doi.org/10.1088/1475-7516/2016/03/018}{JCAP {\normalfont
  \bfseries 1603} (2016) no.~03, 018},
\href{http://arxiv.org/abs/1510.08063}{{\normalfont \ttfamily
  arXiv:1510.08063}}.

\bibitem{Han:1998sg}
T.~Han, J.~D. Lykken, and R.-J. Zhang, ``{\em {On Kaluza-Klein states from
  large extra dimensions}},''
  \href{http://dx.doi.org/10.1103/PhysRevD.59.105006}{Phys. Rev. {\normalfont
  \bfseries D59} (1999)  105006},
\href{http://arxiv.org/abs/hep-ph/9811350}{{\normalfont \ttfamily
  arXiv:hep-ph/9811350}}.

\bibitem{Ackermann:2013yva}
{\normalfont \bfseries Fermi-LAT}, M.~Ackermann {\em et al.}, ``{\em {Dark
  matter constraints from observations of 25 Milky Way satellite galaxies with
  the Fermi Large Area Telescope}},''
  \href{http://dx.doi.org/10.1103/PhysRevD.89.042001}{Phys. Rev. {\normalfont
  \bfseries D89} (2014)  042001},
\href{http://arxiv.org/abs/1310.0828}{{\normalfont \ttfamily arXiv:1310.0828}}.

\bibitem{Fermi-LAT:2016uux}
{\normalfont \bfseries DES, Fermi-LAT}, A.~Albert {\em et al.}, ``{\em
  {Searching for Dark Matter Annihilation in Recently Discovered Milky Way
  Satellites with Fermi-LAT}},''
  \href{http://dx.doi.org/10.3847/1538-4357/834/2/110}{Astrophys. J.
  {\normalfont \bfseries 834} (2017) no.~2, 110},
\href{http://arxiv.org/abs/1611.03184}{{\normalfont \ttfamily
  arXiv:1611.03184}}.

\bibitem{Ade:2015xua}
{\normalfont \bfseries Planck}, P.~A.~R. Ade {\em et al.}, ``{\em {Planck 2015
  results. XIII. Cosmological parameters}},''
  \href{http://dx.doi.org/10.1051/0004-6361/201525830}{Astron. Astrophys.
  {\normalfont \bfseries 594} (2016)  A13},
\href{http://arxiv.org/abs/1502.01589}{{\normalfont \ttfamily
  arXiv:1502.01589}}.

\bibitem{Liu:2016cnk}
H.~Liu, T.~R. Slatyer, and J.~Zavala, ``{\em {Contributions to cosmic
  reionization from dark matter annihilation and decay}},''
  \href{http://dx.doi.org/10.1103/PhysRevD.94.063507}{Phys. Rev. {\normalfont
  \bfseries D94} (2016) no.~6, 063507},
\href{http://arxiv.org/abs/1604.02457}{{\normalfont \ttfamily
  arXiv:1604.02457}}.

\bibitem{Bringmann:2016din}
T.~Bringmann, F.~Kahlhoefer, K.~Schmidt-Hoberg, and P.~Walia, ``{\em {Strong
  constraints on self-interacting dark matter with light mediators}},''
  \href{http://dx.doi.org/10.1103/PhysRevLett.118.141802}{Phys. Rev. Lett.
  {\normalfont \bfseries 118} (2017) no.~14, 141802},
\href{http://arxiv.org/abs/1612.00845}{{\normalfont \ttfamily
  arXiv:1612.00845}}.

\bibitem{Binder:2017lkj}
T.~Binder, M.~Gustafsson, A.~Kamada, S.~M.~R. Sandner, and M.~Wiesner, ``{\em
  {Reannihilation of self-interacting dark matter}},''
  \href{http://dx.doi.org/10.1103/PhysRevD.97.123004}{Phys. Rev. {\normalfont
  \bfseries D97} (2018) no.~12, 123004},
\href{http://arxiv.org/abs/1712.01246}{{\normalfont \ttfamily
  arXiv:1712.01246}}.

\bibitem{Hufnagel:2017dgo}
M.~Hufnagel, K.~Schmidt-Hoberg, and S.~Wild, ``{\em {BBN constraints on
  MeV-scale dark sectors. Part I. Sterile decays}},''
  \href{http://dx.doi.org/10.1088/1475-7516/2018/02/044}{JCAP {\normalfont
  \bfseries 1802} (2018)  044},
\href{http://arxiv.org/abs/1712.03972}{{\normalfont \ttfamily
  arXiv:1712.03972}}.

\bibitem{Hufnagel:2018bjp}
M.~Hufnagel, K.~Schmidt-Hoberg, and S.~Wild, ``{\em {BBN constraints on
  MeV-scale dark sectors. Part II. Electromagnetic decays}},''
\href{http://arxiv.org/abs/1808.09324}{{\normalfont \ttfamily
  arXiv:1808.09324}}.

\bibitem{Essig:2013lka}
R.~Essig {\em et al.}, ``{\em {Working Group Report: New Light Weakly Coupled
  Particles}},'' in {\em {Proceedings, 2013 Community Summer Study on the
  Future of U.S. Particle Physics: Snowmass on the Mississippi (CSS2013):
  Minneapolis, MN, USA, July 29-August 6, 2013}}.
\newblock 2013.
\newblock \href{http://arxiv.org/abs/1311.0029}{{\normalfont \ttfamily
  arXiv:1311.0029}}.
\newblock
\url{http://www.slac.stanford.edu/econf/C1307292/docs/IntensityFrontier/NewLight-17.pdf}.
\newblock

\bibitem{Bernal:2017kxu}
N.~Bernal, M.~Heikinheimo, T.~Tenkanen, K.~Tuominen, and V.~Vaskonen, ``{\em
  {The Dawn of FIMP Dark Matter: A Review of Models and Constraints}},''
  \href{http://dx.doi.org/10.1142/S0217751X1730023X}{Int. J. Mod. Phys.
  {\normalfont \bfseries A32} (2017) no.~27, 1730023},
\href{http://arxiv.org/abs/1706.07442}{{\normalfont \ttfamily
  arXiv:1706.07442}}.

\bibitem{Briceno:2017max}
R.~A. Briceno, J.~J. Dudek, and R.~D. Young, ``{\em {Scattering processes and
  resonances from lattice QCD}},''
  \href{http://dx.doi.org/10.1103/RevModPhys.90.025001}{Rev. Mod. Phys.
  {\normalfont \bfseries 90} (2018) no.~2, 025001},
\href{http://arxiv.org/abs/1706.06223}{{\normalfont \ttfamily
  arXiv:1706.06223}}.

\bibitem{Oman:2015xda}
K.~A. Oman {\em et al.}, ``{\em {The unexpected diversity of dwarf galaxy
  rotation curves}},'' \href{http://dx.doi.org/10.1093/mnras/stv1504}{Mon. Not.
  Roy. Astron. Soc. {\normalfont \bfseries 452} (2015) no.~4, 3650--3665},
\href{http://arxiv.org/abs/1504.01437}{{\normalfont \ttfamily
  arXiv:1504.01437}}.

\bibitem{Kamada:2016euw}
A.~Kamada, M.~Kaplinghat, A.~B. Pace, and H.-B. Yu, ``{\em {How the
  Self-Interacting Dark Matter Model Explains the Diverse Galactic Rotation
  Curves}},'' \href{http://dx.doi.org/10.1103/PhysRevLett.119.111102}{Phys.
  Rev. Lett. {\normalfont \bfseries 119} (2017) no.~11, 111102},
\href{http://arxiv.org/abs/1611.02716}{{\normalfont \ttfamily
  arXiv:1611.02716}}.

\bibitem{Creasey:2016jaq}
P.~Creasey, O.~Sameie, L.~V. Sales, H.-B. Yu, M.~Vogelsberger, and J.~Zavala,
  ``{\em {Spreading out and staying sharp Ð creating diverse rotation curves
  via baryonic and self-interaction effects}},''
  \href{http://dx.doi.org/10.1093/mnras/stx522}{Mon. Not. Roy. Astron. Soc.
  {\normalfont \bfseries 468} (2017) no.~2, 2283--2295},
\href{http://arxiv.org/abs/1612.03903}{{\normalfont \ttfamily
  arXiv:1612.03903}}.

\bibitem{Robertson:2017mgj}
A.~Robertson {\em et al.}, ``{\em {The diverse density profiles of galaxy
  clusters with self-interacting dark matter plus baryons}},''
  \href{http://dx.doi.org/10.1093/mnrasl/sly024}{Mon. Not. Roy. Astron. Soc.
  {\normalfont \bfseries 476} (2018) no.~1, L20--L24},
\href{http://arxiv.org/abs/1711.09096}{{\normalfont \ttfamily
  arXiv:1711.09096}}.

\bibitem{Bethe:1949yr}
H.~A. Bethe, ``{\em {Theory of the Effective Range in Nuclear Scattering}},''
\href{http://dx.doi.org/10.1103/PhysRev.76.38}{Phys. Rev. {\normalfont
  \bfseries 76} (1949)  38--50}.

\end{thebibliography}\endgroup
\appendix 
\section{Supplementary Material}


Here we solve Eq.~(5) of the main text  for $v_\text{max} \to \infty$. In the narrow width approximation, \textit{i.e.} when the second term in the denominator is much smaller than the first one,  we can do the replacement 
\begin{equation}
\frac{1}{(v^2-v_R^2)^2+16\gamma^2 v^{2(2L+1)}}\to \frac{\pi \delta(v-v_R)}{8\gamma v_R^{2(L+1)}}\,.
\label{eq:swa}
\end{equation}
This leads to 
\begin{equation}
{\cal I}_L (\gamma, v_R,v_0) \simeq \frac{\pi\gamma}{8} f(v_R,v_0) v_R^{2L-1}   \,,
\label{eq:swad}
\end{equation}
which is  valid for $\gamma \ll v_R^{1-2L}$.  
If  $\gamma \lesssim{\cal O}( 1)$, which is the region of interest in this work,  this approximation is very good for any $L\neq 0$. We have corroborated this numerically. For $L=0$, such an approximation does not always work. That is however not a problem because there is an exact formula in terms of  the  exponential integral function, Ei$(z)$, defined by the principal value of  $-\int^\infty_{-z} e^{-t}dt/t$. 

For this, let us first notice  that when $L=0$ one can rearrange the denominator of Eq.~(5) of the main text in terms of the integrals
\begin{eqnarray}
{\cal I}_\pm (\gamma, v_R,v_0) &\equiv&
\int^\infty_0  \frac{\gamma^2 f(v,v_0) v\, dv}{(v^2\mp v_R^2)^2\pm 16\gamma^2}\,,
\end{eqnarray}
which are defined for $\gamma$ and $v_R$ real.  Specifically
\begin{align}
\lefteqn{
{\cal I}_{L=0} (\gamma, v_R,v_0) =}\\
&\frac{1}{\pm(v_R^2 -4 \gamma^2)}{\cal I}_\pm \left(\gamma  \sqrt{\pm (v_R^2-4\gamma^2)},\sqrt{\pm (v_R^2-8\gamma^2)},v_0 \right)\,,\nonumber
\end{align}
where the plus sign applies for $\gamma^2<v_R^2/8$ and the negative one for $\gamma^2>v_R^2/4$.  (For simplicity, we do not report the expression for the narrow range $v_R^2/8<\gamma^2 <v_R^2/4$). Finally, ${\cal I}_- $ can be calculated by means of
\begin{align}
\lefteqn{ {\cal I}_- (\gamma, v_R,v_0) =} \\
&\frac{ \gamma }{16 } f \left( i \sqrt{v_R^2+4\gamma} ,v_0\right) {\rm Ei}\left(-\frac{v_R^2+4\gamma}{v_0^2}\right) 
+(\gamma\to-\gamma)\,. \nonumber
\end{align}
Moreover, by analytically extending the previous expression, we can calculate the other integral. This is
\begin{align}
{\cal I}_+ (\gamma, v_R,v_0) 
= &- {\cal I}_-( i \gamma, i v_R, v_0 )
\\+\frac{\pi\gamma }{8 v_R^2} f(v_R,v_0)&\left( v_R^2\cos \left(\frac{4\gamma}{v_0^2}\right)+4\gamma\sin \left(\frac{4\gamma}{v_0^2}\right)\right)\nonumber
\,.
\end{align}
As a check, one can take the limit $\gamma\to0$ and recover the narrow width approximation of Eq.~\eqref{eq:swad}.

\end{document}